\documentclass[
 showpacs,preprintnumbers,
 amsmath,amssymb,
prb,
twocolumn]{revtex4-1}

\usepackage{mathrsfs}
\usepackage{graphicx}
\usepackage{subfigure}
\usepackage{dcolumn}
\usepackage{bm}
\usepackage{bbm}
\usepackage{color}
\usepackage{esint}
\usepackage{lineno}

\begin{document}


\title{Majorana transport in superconducting nanowire with Rashba and Dresselhaus spin-orbit couplings}

\author{Jia-Bin You$^{1,}$}
\email{jiabinyou@gmail.com}

\author{Xiao-Qiang Shao$^{1,3}$}

\author{Qing-Jun Tong$^{1}$}

\author{A. H. Chan$^{2,}$}
\email{phycahp@nus.edu.sg}

\author{C. H. Oh$^{1,2,}$}
\email{phyohch@nus.edu.sg}

\author{Vlatko Vedral$^{1,2,4,}$}
\email{phyvv@nus.edu.sg}

\affiliation{$^1$Centre for Quantum Technologies, National
University of Singapore, 117543, Singapore\\$^2$Department of
Physics, National University of Singapore, 117542,
Singapore\\$^3$School of Physics, Northeast Normal University, Changchun, 130024, People's Republic of China\\$^4$Department of Physics, University of Oxford, Clarendon Laboratory, Oxford, OX1 3PU, United Kingdom}

\date{\today}

\begin{abstract}

Tunneling experiment is a key technique for detecting Majorana fermion in solid state systems. We use Keldysh non-equilibrium Green function method to study multi-lead tunneling in superconducting nanowire with Rashba and Dresselhaus spin-orbit couplings. A zero-bias \textit{dc} conductance peak appears in our setup which signifies the existence of Majorana fermion and is in accordance with previous experimental results on InSb nanowire. Interestingly, due to the exotic property of Majorana fermion, there exists a hole transmission channel which makes the currents asymmetric at the left and right leads.
The \textit{ac} current response mediated by Majorana fermion is also studied here. To discuss the impacts of Coulomb interaction and disorder on the transport property of Majorana nanowire, we use the renormalization group method to study the phase diagram of the wire. It is found that there is a topological phase transition under the interplay of superconductivity and disorder. We find that the Majorana transport is preserved in the superconducting-dominated topological phase and destroyed in the disorder-dominated non-topological insulator phase.

\end{abstract}

\maketitle


\section{Introduction}

An intensive search is ongoing in experimental realization of topological superconductor for topological quantum computing \cite{PhysRevLett.103.020401,PhysRevB.82.134521,PhysRevB.87.054501,Jiabin2013,PhysRevB.88.060504,PhysRevLett.104.040502,PhysRevB.81.125318,PhysRevB.82.184525,PhysRevB.61.10267}. The basic idea is to embeds qubit in a nonlocal, intrinsically decoherence-free way. The prototype is a spinless $p$-wave superconductor \cite{Kitaev2001,Kitaev20032,PhysRevLett.86.268}. Edge excitations in such a state are Majorana fermions (MFs) which obey non-Abelian statistics and can be manipulated by braiding operations. The nonlocal MFs are robust against local perturbations and have been proposed for topological quantum information processing \cite{Alicea2011,PhysRevB.84.035120}.

\begin{figure}
\includegraphics[width=7cm]{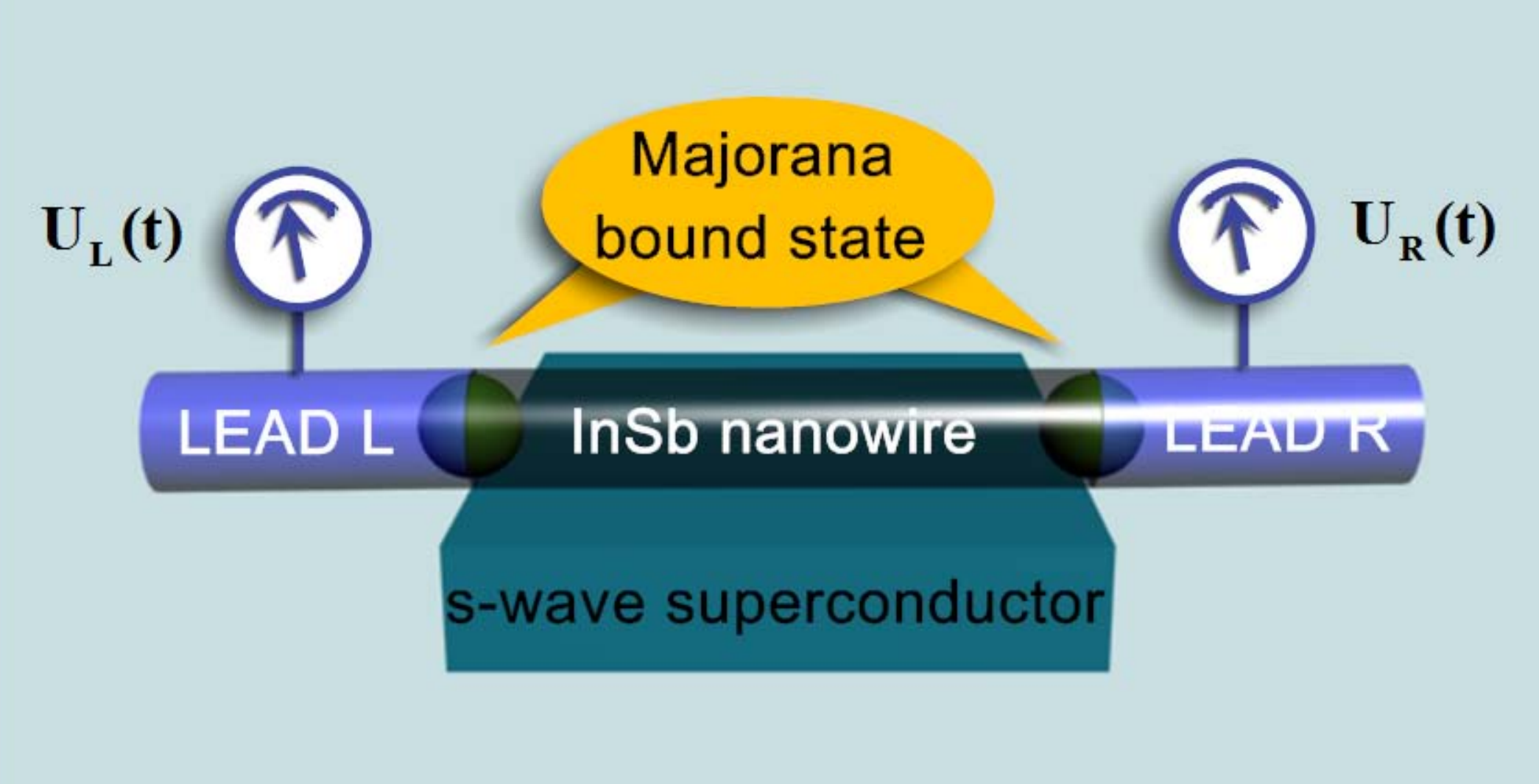}
\caption{(color online). Experimental setup for tunneling
experiment. An InSb nanowire is deposited on an $s$-wave
superconductor and coupled to two normal metal leads.}\label{illustration}
\end{figure}

A hybrid semiconducting-superconducting nanostructure has become a mainstream experimental setup recently for realizing topological superconductor and Majorana fermion \cite{PhysRevLett.100.096407,PhysRevLett.104.040502,PhysRevB.81.125318,PhysRevLett.105.177002}. The signature of MFs characterized by a zero-bias conductance peak (ZBP) has been reported in the tunneling experiments of the InSb nanowire \cite{Mourik2012,NatPhys.8.887,NatNano.9.79,PhysRevB.87.241401,PhysRevLett.110.126406}. Motivated by this, we propose a multi-lead setup for studying the tunneling transport of MFs as shown in Fig. \ref{illustration}. A spin-orbit coupled InSb nanowire is deposited on an $s$-wave superconductor. Due to the
superconducting proximity effect, the wire is effectively equivalent to the spinless $p$-wave superconductor and hosts MFs at the ends. The nanowire is then coupled to two normal metal leads so as to measure the currents. For our study, we apply the Keldysh non-equilibrium Green function (NEGF) method to obtain the current response of the tunneling Hamiltonian \cite{PhysRevLett.68.2512,PhysRevB.84.165440,PhysRevB.50.5528,PhysRevB.82.180516,Datta2005,PhysRevB.88.024511,PhysRevLett.111.136402}. Curiously in the multi-lead case, we observe that the currents at left and right leads are asymmetric as shown in Fig. \ref{cas}. This is due to the exotic commutation relation of MFs, $\{\gamma_{i},\gamma_{j}\}=2\delta_{i,j}$. From another standpoint, the zero-energy fermion $b_{0}$ combined by the end-Majorana modes ($\gamma_{L,R}$) is so highly nonlocal, $b_{0}=(\gamma_{L}+i\gamma_{R})/2$, as to make the Majorana transport deviate from the ordinary transport mediated by electron. Different from the ordinary one, there is a new transmission channel (hole-channel in Eq. (\ref{current3})) in Majorana transport. This makes the left and right currents asymmetric. The current asymmetry may be used as a criterion to further confirm the existence of the Majorana fermion.
We also give the \textit{ac} current response in our work and find that the current is enhanced in step with the increase of level broadening and the decrease of temperature, and finally saturates at high voltage. We use the bosonization and renormalization group (RG) methods to consider the transport property of the Majorana nanowire with short-range Coulomb interaction and disorder \cite{Giamarchi,PhysRevB.37.325,PhysRevB.53.3211,PhysRevB.87.060504,PhysRevB.84.014503,PhysRevB.78.054436,PhysRevLett.109.146403,PhysRevLett.109.166403}. We observe that there is a topological quantum phase transition under the interplay of superconductivity and disorder. It is found that the Majorana transport is preserved in the superconducting-dominated topological phase and destroyed in the disorder-dominated non-topological insulator phase. The phase diagram and the condition in which the Majorana transport exists are given.

\section{Model}\label{model}

The model is depicted in Fig. \ref{illustration}. Two normal metal leads are connected to the superconducting wire through ohmic contacts at the two ends. When the chemical potential of superconducting wire lies within the energy gap, two MFs will appear at the two
ends of the wire, respectively. The topological superconducting wire is made of a spin-orbit coupled semiconductor (InSb wire) depositing on an $s$-wave superconducting substrate. Via the superconducting proximity effect \cite{PhysRevLett.100.096407}, the Cooper pair will tunnel into the semiconductor and generate the $s$-wave superconductivity in the semiconducting wire.

The one dimensional spin-orbit coupled $s$-wave superconducting nanowire can be modeled as $H_{\text{nw}}=H_0+H_{\Delta}$ \cite{PhysRevB.84.014503,PhysRevB.87.235414}, where
\begin{equation}
\label{H0}
\begin{split}
&H_{0}=\int dk \mathbf{\Psi}_{k}^{\dag}[\xi_{k}+(\alpha\sigma_{y}+\beta\sigma_{x})k+V_{z}\sigma_{z}]\mathbf{\Psi}_{k},\\
&H_{\Delta}=\Delta\int dk (a_{k\uparrow}a_{-k\downarrow}+\text{H.c.}).
\end{split}
\end{equation}
Here $\xi_{k}=k^2/2m-\mu$ where $k$ is the momentum and $\mu$ is the chemical potential, $\sigma_{x}$ and $\sigma_{y}$ are spin Pauli matrices,
$\alpha$ and $\beta$ are the Rashba and Dresselhaus spin-orbit
strengths, $\Delta$ is the $s$-wave gap function and $\mathbf{\Psi}_{k}=(a_{k\uparrow},a_{k\downarrow})^{T}$ where $a_{k\uparrow}$ ($a_{k\downarrow}$) is the annihilation operator for spin up (down) electron. We have exerted a perpendicular magnetic field $V_{z}$ on the wire and considered the Zeeman effect.

In the Nambu basis $\mathbf{\Phi}_{k}^{\dag}=(\mathbf{\Psi}_{k}^{\dag},\mathbf{\Psi}_{-k}^{T})$, the Hamiltonian Eq. (\ref{H0}) can be recast into $H=\frac{1}{2}\int dk\mathbf{\Phi}_{k}^{\dag}\mathcal{H}(k)\mathbf{\Phi}_{k}$, where
\begin{equation}
\label{topowireMS}
\begin{split}
\mathcal{H}(k)=\xi_{k}\sigma_{z}+\alpha k \sigma_{z}\tau_{y}+\beta k\tau_{x}+\Delta\sigma_{y}\tau_{y}+V_{z}\sigma_{z}\tau_{z}.\\
\end{split}
\end{equation}
Here $\tau_{x}$, $\tau_{y}$ and $\tau_{z}$ are the Pauli matrices in the particle-hole space. It is known that the BdG Hamiltonian Eq. (\ref{topowireMS}) satisfies particle-hole symmetry, $\Xi^{-1}\mathcal{H}(k)\Xi=-\mathcal{H}(-k)$, where $\Xi=\sigma_{x}K$ and $K$ is the complex conjugation operator \cite{PhysRevB.87.054501,Jiabin2013}.
The topological property of this BdG Hamiltonian can be examined by the Pfaffian invariant,
\begin{equation}
\label{pfaffian}
\begin{split}
\mathcal{P}=\text{sgn}\left\{{\text{Pf}[\mathcal{H}(k=0)\sigma_{x}]}\right\}=\text{sgn}(\mu^2+\Delta^2-V_{z}^2).\\
\end{split}
\end{equation}
Therefore, a topological quantum phase transition occurs when $\mu^2+\Delta^2=V_{z}^2$. For $\mu^2+\Delta^2<V_{z}^2$, $\mathcal{P}=-1$, the gap is dominated by the magnetic field and the wire is in the topological phase with Majorana fermion at the ends of the nanowire. For $\mu^2+\Delta^2>V_{z}^2$, $\mathcal{P}=1$, the gap is dominated by pairing with no end states. In this work, we study the case where the nanowire is in the topological phase. This can be realized by putting the chemical potential inside the energy gap. The low energy theory of the Hamiltonian Eq. (\ref{H0}) can then be obtained as follow. Diagonalizing the
Hamiltonian $H_0$, we get two energy bands, $\varepsilon_{\pm}(k)=\frac{k^2}{2m}-\mu\pm\sqrt{(\alpha^2+\beta^2)k^2+V_{z}^{2}}$.
For these two bands, the eigenstates are
\begin{equation}
\begin{split}
|\chi_{+}(k)\rangle=\left[\begin{array}{*{20}c}
{e^{-i\theta/2}\cos\frac{\gamma_{k}}{2}}\\
{e^{i\theta/2}\sin\frac{\gamma_{k}}{2}}\\
\end{array}\right], |\chi_{-}(k)\rangle=\left[\begin{array}{*{20}c}
{-e^{-i\theta/2}\sin\frac{\gamma_{k}}{2}}\\
{e^{i\theta/2}\cos\frac{\gamma_{k}}{2}}\\
\end{array}\right],
\end{split}
\end{equation}
respectively, where $\tan\theta=\alpha/\beta$ and
$\tan\gamma_{k}=\sqrt{\alpha^2+\beta^2}k/V_{z}$. When the magnetic field is  dominant than the spin-orbit interactions ($V_{z}\gg\alpha,\beta$), the spins will be forced to be nearly polarized within each band. Because the chemical
potential lies within the gap, only the low energy band is
near the Fermi points and activated. We can thus restrict the Hilbert space to the lower band in this case. To achieve this, we unitarily transform the electron operator from spin basis to band basis, $(a_{k+}^{\dag},a_{k-}^{\dag})=(a_{k\uparrow}^{\dag},a_{k\downarrow}^{\dag})U$, where $U=(|\chi_{+}(k)\rangle,|\chi_{-}(k)\rangle)$. Here $a_{k+}^{\dag}$ ($a_{k-}^{\dag}$) is the creation operator for upper (lower) band. Then we neglect the upper band and obtain the low energy approximation of the Hamiltonian $H_{0}=\int dk\varepsilon_{-}(k)d_{k}^{\dag}d_{k}$, where $d_{k}\equiv a_{k-}$.
Similarly, projecting the superconducting term onto the lower band
$|\chi_{-}(k)\rangle$, we have $H_{\Delta}=-\frac{\Delta}{2}\int
dk (\sin{\gamma_{k}}d_{k}d_{-k}+\text{H.c.})$.

Therefore, the low energy theory for the topological superconductivity in the spin-orbit coupled
semiconducting nanowire deposited on an $s$-wave superconductor is exhibited by $H_{\text{nw}}=\int dk(k^2/2m-\mu_{\text{eff}})d_{k}^{\dag}d_{k}-\Delta_{\text{eff}}(kd_{k}d_{-k}+\text{H.c.})$,
where $\mu_{\text{eff}}=\mu+|V_{z}|$ and
$\Delta_{\text{eff}}=\frac{\Delta\sqrt{\alpha^2+\beta^2}}{2|V_{z}|}$.
The Hamiltonian $H_{\text{nw}}$ is exactly the spinless $p$-wave superconductor and
has been shown that \cite{Kitaev2001} there exist unpaired Majorana fermions at the
left and right end sides of the nanowire. The effective Hamiltonian for this piece of the system is
\begin{equation}
\begin{split}
H_{\text{mf}}=\frac{i}{2}t(\gamma_{L}\gamma_{R}-\gamma_{R}\gamma_{L}),\\
\end{split}
\end{equation}
where $\gamma_{L/R}$ is the Majorana operator at the left/right end side and $t\sim e^{-L/l_{0}}$ describes the coupling energy between the two
MFs, $L$ is the length of wire, and $l_{0}$ is the superconducting coherence length.

We next focus on the tunneling transport of Majorana nanowire described by $H_{\text{mf}}$. Guided by the typical experimental setup in which the leads are made of gold, we view electrons in the leads as noninteracting. We then apply time-dependent bias voltages on the left and right leads respectively. This can be physically described by
\begin{equation}
\begin{split}
H_{s}=\sum_{p}\xi_{p,s}(t)c_{p,s}^{\dag}c_{p,s},\\
\end{split}
\end{equation}
where $s=L,R$, and $c_{p,s}$ is the electron annihilation operator for the lead. Here $\xi_{p,s}(t)=\varepsilon_{p,s}-eU_{s}(t)$, where $\varepsilon_{p,s}$ is the dispersion relation for the metallic lead and $U_{s}(t)$ is the time-dependent bias voltage on the lead. Note that the occupation for each lead is determined by the equilibrium distribution function established before the time-dependent bias voltage and tunneling are turned on.
The tunneling between the leads and the wire is dependent upon the geometry of experimental layout and upon the self-consistent response of charge in the leads to the time-dependent bias voltages \cite{PhysRevB.50.5528}. We can simply express the tunneling as
\begin{equation}
\begin{split}
H_{T,s}=\sum_{pi}[V_{pi,s}^{*}(t)c_{p,s}^{\dag}-V_{pi,s}(t)c_{p,s}]\gamma_{i},\\
\end{split}
\end{equation}
where $i,s=L,R$, and $V_{pi,s}(t)$ is the tunneling strength.
Therefore, the Hamiltonian for the experimental setup of Fig. \ref{illustration} can be described by $H=H_{L}+H_{TL}+H_{\text{mf}}+H_{TR}+H_{R}$.

\section{NEGF method for the Majorana Current}

The Keldysh nonequilibrium Green function technique is used very widely to describe transport phenomena in mesoscopic systems. In the tunneling problem formulated in Sec. \ref{model}, we consider the time-dependent bias voltages and tunneling strengths. This is essentially a nonequilibrium problem and can be treated by the Keldysh formalism. In this formalism, the leads and the wire are decoupled and each part is in thermal equilibrium characterized by their respective chemical potentials at $t=-\infty$. We first adiabatically evolve the system by the total Hamiltonian $H$ from $t=-\infty$ to $t=+\infty$, then evolve the system back in time from $t=+\infty$ to $t=-\infty$, and calculate the physical quantity during this evolution. Finally the system is back in the initial state at $t=-\infty$. This procedure eliminates the uncertain state at the asymptotically large time in the nonequilibrium theory. The time loop, which contains two pieces: the outgoing branch from $t=-\infty$ to $t=+\infty$ and the ingoing branch from $t=+\infty$ to $t=-\infty$, is called Keldysh contour. Below we will use the Keldysh NEGF method to study the Majorana current in the tunneling transport.

\subsection{general formula}

We study the Majorana current from the
left/right lead to the wire. The current is given by the changing rate of charge in the lead, $I_{s}=-e\langle \dot{N}_{s}\rangle$, where $s=L,R$, $N_{s}$ is the number operator in the lead, $N_{s}=\sum_{p}c_{p,s}^{\dag}c_{p,s}$. The bracket $\langle\rangle$ denotes the ensemble average with respect to the Hamiltonian of experimental setup $H$. The commutation relations of electrons and MFs are $\{c_{p,s},c_{p',s'}^{\dag}\}=\delta_{p,p'}\delta_{s,s'}$ and $\{\gamma_{i},\gamma_{j}\}=2\delta_{i,j}$, and zero otherwise. Using the Heisenberg equation, the current from the lead to the wire is
\begin{equation}
\label{current1}
\begin{split}
I_{s}(t)&=-e\langle \dot{N}_{s}\rangle=-ie\langle[H,N_{s}]\rangle=-ie\langle[H_{T,s},N_{s}]\rangle,\\
&=ie\sum_{pi}\langle V_{pi,s}^{*}(t)c_{p,s}^{\dag}\gamma_{i}-V_{pi,s}(t)\gamma_{i}c_{p,s}\rangle,\\
&=2e\sum_{pi}\text{Re}\{V_{pi,s}^{*}(t)\langle ic_{p,s}^{\dag}\gamma_{i}\rangle\},\\
&=2e\sum_{pi}\text{Re}\{V_{pi,s}^{*}(t)G_{ip,s}^{<}(t,t)\},\\
\end{split}
\end{equation}
where $G_{ip,s}^{<}(t,t')=i\langle c_{p,s}^{\dag}(t')\gamma_{i}(t)\rangle$ is the lesser component of the Keldysh Green function
\begin{equation}
\begin{split}
G_{ip,s}(t,t')=-i\langle T_{K}\gamma_{i}(t)c_{p,s}^{\dag}(t')\rangle.\\
\end{split}
\end{equation}
Here operator $T_{K}$ orders the times along the Keldysh contour with earlier times occurring first.

To proceed, we express the coupling Green function $G_{ip}$ as a product of Green functions for the lead $G_{p}$ and the wire $G_{ij}$. Via the equation of motion (EOM) method \cite{Mahan1990}, we have
\begin{equation}
\label{cpG}
\begin{split}
G_{ip,s}(t,t')=\sum_{j}\int_{K} dt''G_{ij}(t,t'')V_{pj,s}(t'')G_{p,s}^{0}(t'',t'),\\
\end{split}
\end{equation}
where
\begin{equation}
\begin{split}
&G_{ij}(t,t')=-i\langle T_{K}\gamma_{i}(t)\gamma_{j}(t')\rangle,\\
&G_{p,s}^{0}(t,t')=-i\langle T_{K}c_{p,s}(t)c_{p,s}^{\dag}(t')\rangle_{0}\\
\end{split}
\end{equation}
are the Green function of the wire and the free Green function of the lead respectively. Here $\langle\rangle_0$ is the ensemble average with respect to the Hamiltonian of lead $H_s$. The integration is taken on the Keldysh contour. Therefore, via the Keldysh Green function method \cite{Mahan1990}, we can get the lesser component of the coupling Green function Eq. (\ref{cpG}) by analytical continuation,
\begin{equation}
\begin{split}
G_{ip,s}^{<}(t,t')=&\sum_{j}\int_{-\infty}^{\infty}dt''V_{pj,s}(t'')[G_{ij}^{R}(t,t'')G_{p,s}^{0<}(t'',t')\\
&+G_{ij}^{<}(t,t'')G_{p,s}^{0A}(t'',t')],\\
\end{split}
\end{equation}
where $G^{R}$ and $G^{A}$ are the retarded and advanced Green functions. The expressions for the free Green functions $G^{0}$ can be found in the Appendix \ref{freeGF}. Substituting this lesser Green function into the current formula Eq. (\ref{current1}) and using the expressions for the free Green functions, we arrive at
\begin{equation}
\label{current2}
\begin{split}
I_{s}(t)=&-2e\text{Im}\Big\{\sum_{ij}\int_{-\infty}^{t}dt_{1}\int_{-\infty}^{\infty}\frac{d\varepsilon}{2\pi}e^{i\varepsilon(t-t_{1})}\times\\
&[\Gamma_{s}(\varepsilon,t_{1},t)]_{ji}[G_{ij}^{R}(t,t_{1})f_{s}(\varepsilon)+G_{ij}^{<}(t,t_{1})]\Big\},\\
\end{split}
\end{equation}
where $f_{s}(\varepsilon)$ is the Fermi function. The time-dependent level broadening matrix is given by
\begin{equation}
\begin{split}
[\Gamma_{s}(\varepsilon,t_{1},t)]_{ji}=2\pi\rho(\varepsilon)V_{i,s}^{*}(\varepsilon,t)V_{j,s}(\varepsilon,t_{1})e^{-ie\int_{t_{1}}^{t}dt_{2}U_{s}(t_{2})},\\
\end{split}
\end{equation}
where the density operator is $\rho(\varepsilon)=\sum_{p}\delta(\varepsilon-\varepsilon_{p,s})$. Here we have explicitly indicated the energy dependence of the tunneling strength $V_{pi,s}(t)$. It is easy to check that the broadening matrix is Hermitian, $\mathbf{\Gamma}_{s}^{\dag}(\varepsilon,t_{1},t)=\mathbf{\Gamma}_{s}(\varepsilon,t,t_{1})$.

For the tunneling strength, by the wide-band approximation \cite{PhysRevB.50.5528}, the momentum and time dependence can be factorized, $V_{pi,s}(t)=V_{i,s}(\varepsilon_{p,s},t)=u_{s}(t)V_{i,s}(\varepsilon_{p,s})$. Thus we find that $[\Gamma_{s}(t)]_{ji}\equiv[\Gamma_{s}(\varepsilon,t,t)]_{ji}=[\Gamma_{s}(\varepsilon)]_{ji}|u_{s}(t)|^2$, where the level broadening matrix is
\begin{equation}
\begin{split}
[\Gamma_{s}(\varepsilon)]_{ji}=2\pi\rho_{s}(\varepsilon)V_{i,s}^{*}(\varepsilon)V_{j,s}(\varepsilon). \end{split}
\end{equation}
Below we assume that the tunneling strength is time-independent and set $u_{s}(t)=1$. In the mesoscopic transport, the physical property is generally dominated by states near the Fermi level. Since the broadening matrix is usually slowly varying function of energy close to the Fermi level, we can assume that it is energy independent, $\mathbf{\Gamma}_{s}(\varepsilon)=\mathbf{\Gamma}_{s}$. This wide-band approximation captures the main physics of the tunneling problem and can be used to simplify the current expression Eq. (\ref{current2}).

Therefore, the current can be further reduced to $I_{s}(t)=I_{s}^{\text{out}}(t)+I_{s}^{\text{in}}(t)$, where
\begin{equation}
\begin{split}
I_{s}^{\text{out}}(t)=-e\text{ImTr}[&\mathbf{\Gamma}_{s}\mathbf{G}^{<}(t,t)],\\
I_{s}^{\text{in}}(t)=-e\text{ImTr}\Big\{&\int_{-\infty}^{\infty}\frac{d\varepsilon}{\pi}f_{s}(\varepsilon)\int_{-\infty}^{t}dt_{1}e^{-i\varepsilon(t_{1}-t)}\times\\
&\mathbf{\Gamma}_{s}(t_{1},t)\mathbf{G}^{R}(t,t_{1})\Big\}.\\
\end{split}
\end{equation}
Here $\mathbf{\Gamma}_{s}(t_{1},t)\equiv\mathbf{\Gamma}_{s}(\varepsilon,t_{1},t)=\mathbf{\Gamma}_{s}e^{-ie\int_{t_{1}}^{t}dt_{2}U_{s}(t_{2})}$ and $\mathbf{G}^{<,R}$ are the Green functions of the wire. The current has been separated in two parts: the outflux, $I_{s}^{\text{out}}(t)$, which is easy to be identified since $\mathbf{\Gamma}_{s}$ represents the rate at which an electron placed initially in the energy level of the wire will escape into the lead and $N(t)=\text{ImTr}[\mathbf{G}^{<}(t,t)]$ is the number of particles in the wire; the influx, $I_{s}^{\text{in}}(t)$, which is proportional to the occupation $f_{s}(\varepsilon)$ in the lead and to the density of states $\rho(\varepsilon)=\text{ImTr}[\mathbf{G}^{R}(\varepsilon)]$ in the wire \cite{Datta2005}.
For the outflow, the lesser Green function can be calculated by the relation
\begin{equation}
\begin{split}
\mathbf{G}^{<}(t,t)&=\int dt_{1}dt_{2}\mathbf{G}^{R}(t,t_{1})\mathbf{\Sigma}^{<}(t_{1},t_{2})\mathbf{G}^{A}(t_{2},t),\\
\end{split}
\end{equation}
where the explicit expression for the lesser self-energy $\mathbf{\Sigma}^{<}(t_{1},t_{2})$ is given in the Appendix \ref{wireGF}. Substituting $\mathbf{\Sigma}^{<}$ into the lesser Green function, we have
\begin{equation}
\begin{split}
\mathbf{G}^{<}(t,t)&=\sum_{s=L,R}\int\frac{d\varepsilon}{2\pi}[if_{s}(\varepsilon)]\times\\
&[\mathbf{A}_{s}(\varepsilon,t)\mathbf{\Gamma}_{s}\mathbf{A}_{s}^{\dag}(\varepsilon,t)+\mathbf{B}_{s}(\varepsilon,t)\mathbf{\Gamma}_{s}^{*}\mathbf{B}_{s}^{\dag}(\varepsilon,t)],\\
\end{split}
\end{equation}
where
\begin{equation}
\label{AB}
\begin{split}
\mathbf{A}_{s}(\varepsilon,t)&=\int dt_{1}e^{-i\varepsilon(t_{1}-t)}e^{ie\int_{t}^{t_{1}}dt'U_{s}(t')}\mathbf{G}^{R}(t,t_{1}),\\
\mathbf{B}_{s}(\varepsilon,t)&=\int dt_{1}e^{-i\varepsilon(t_{1}-t)}e^{-ie\int_{t}^{t_{1}}dt'U_{s}(t')}\mathbf{G}^{R}(t,t_{1}).\\
\end{split}
\end{equation}
Since $\text{Tr}[\mathbf{\Gamma}_{s}\mathbf{A}\mathbf{\Gamma}_{s'}\mathbf{A}^{\dag}]$ is real, the outflow can be finally written as
\begin{equation}
\label{outflow}
\begin{split}
I_{s}^{\text{out}}(t)&=-e\sum_{s'=L,R}\int\frac{d\varepsilon}{2\pi}f_{s'}(\varepsilon)\text{Tr}\{\mathbf{\Gamma}_{s}\mathbf{A}_{s'}(\varepsilon,t)\mathbf{\Gamma}_{s'}\mathbf{A}_{s'}^{\dag}(\varepsilon,t)\\
&+\mathbf{\Gamma}_{s}\mathbf{B}_{s'}(\varepsilon,t)\mathbf{\Gamma}_{s'}^{*}\mathbf{B}_{s'}^{\dag}(\varepsilon,t)\}.\\
\end{split}
\end{equation}
For the inflow, after some calculations, we obtain
\begin{equation}
\label{inflow}
\begin{split}
I_{s}^{\text{in}}(t)&=-e\int\frac{d\varepsilon}{\pi}f_{s}(\varepsilon)\text{ImTr}\{\mathbf{\Gamma}_{s}\mathbf{A}_{s}(\varepsilon,t)\}.\\
\end{split}
\end{equation}
The retarded Green function of the wire is deduced in the Appendix \ref{wireGF}. Here we only show the result,
\begin{equation}
\begin{split}
\mathbf{G}^{R}(t,t_{1})=-2i\theta(t-t_{1})e^{(2\mathbf{t}-\mathbf{\Gamma})(t-t_{1})},
\end{split}
\end{equation}
where $\theta(t)$ is the step function. The broadening matrix and the hopping matrix are
\begin{equation}
\begin{split}
\mathbf{\Gamma}=\left[\begin{array}{*{20}c}
{2\Gamma_{L}} & {0}\\
{0} & {2\Gamma_{R}}\\
\end{array}\right], \mathbf{t}=\left[\begin{array}{*{20}c}
{0} & {t}\\
{-t} & {0}\\
\end{array}\right],
\end{split}
\end{equation}
where $\Gamma_{L}$ and $\Gamma_{R}$ are positive.

\subsection{\textit{dc} current response}

\begin{figure}
\hspace{-0.04\textwidth}\includegraphics[width=6cm]{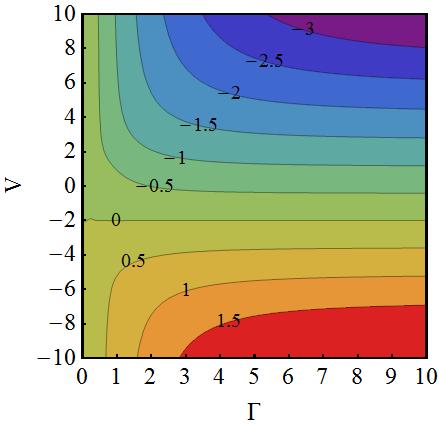}
\caption{(color online). Current asymmetry $J=I_{L}+I_{R}$ in the $\Gamma\sim V$ plane ranging from $-3$ to $+1.5$. We set $U_{L}=1$ and $\Gamma_{L}=\Gamma_{R}=\Gamma$ in this diagram.}\label{cas}
\end{figure}

We first discuss the current response to the \textit{dc} voltages at the two ends, $U_{L}(t)=U_{L}$ and $U_{R}(t)=U_{R}$. For the homogeneous system, $\mathbf{A}_{s}(\varepsilon,t)$ and
$\mathbf{B}_{s}(\varepsilon,t)$ are time-independent and just the Fourier transform of the retarded Green function, $\mathbf{A}_{s}(\varepsilon,t)=\mathbf{G}^{R}(\varepsilon-eU_{s})$ and $\mathbf{B}_{s}(\varepsilon,t)=\mathbf{G}^{R}(\varepsilon+eU_{s})$. Substituting this relation into the outflow Eq. (\ref{outflow}), we immediately have
\begin{equation}
\begin{split}
I_{s}^{\text{out}}=-e\sum_{s'}\int&\frac{d\varepsilon}{2\pi}[T_{ss'}^{e}(\varepsilon)f_{s'}(\varepsilon+eU_{s'})\\
&+T_{ss'}^{h}(\varepsilon)f_{s'}(\varepsilon-eU_{s'})],\\
\end{split}
\end{equation}
where the particle and hole transmission functions are
$T_{ss'}^{e}(\varepsilon)=\text{Tr}[\mathbf{\Gamma}_{s}\mathbf{G}^{R}(\varepsilon)\mathbf{\Gamma}_{s'}\mathbf{G}^{A}(\varepsilon)]$
and
$T_{ss'}^{h}(\varepsilon)=\text{Tr}[\mathbf{\Gamma}_{s}\mathbf{G}^{R}(\varepsilon)\mathbf{\Gamma}_{s'}^{*}\mathbf{G}^{A}(\varepsilon)]$,
respectively. Similarly, the inflow can be written as
\begin{equation}
\label{inflow2}
\begin{split}
I_{s}^{\text{in}}&=-e\int\frac{d\varepsilon}{\pi}f_{s}(\varepsilon+eU_{s})\text{ImTr}\{\mathbf{\Gamma}_{s}\mathbf{G}^{R}(\varepsilon)\},\\
&=-e\int\frac{d\varepsilon}{2\pi i}f_{s}(\varepsilon+eU_{s})\text{Tr}\{\mathbf{\Gamma}_{s}[\mathbf{G}^{R}(\varepsilon)-\mathbf{G}^{A}(\varepsilon)]\}.\\
\end{split}
\end{equation}
Here we have used the fact that the retarded and advanced Green functions are conjugated. By the definitions of the retarded and advanced Green functions, we have $\mathbf{G}^{R}(\varepsilon)-\mathbf{G}^{A}(\varepsilon)=-i\sum_{s'}\mathbf{G}^{R}(\varepsilon)(\Gamma_{s'}+\Gamma_{s'}^{*})\mathbf{G}^{A}(\varepsilon)$ \cite{Mahan1990}. Substituting this relation into the inflow Eq. (\ref{inflow2}), we have
\begin{equation}
\begin{split}
I_{s}^{\text{in}}=e\sum_{s'}\int&\frac{d\varepsilon}{2\pi}[T_{ss'}^{e}(\varepsilon)f_{s}(\varepsilon+eU_{s})\\
&+T_{ss'}^{h}(\varepsilon)f_{s}(\varepsilon+eU_{s})].\\
\end{split}
\end{equation}
Therefore, for the \textit{dc} case, the current is reduced to the
Landauer-B\"{u}ttiker formula \cite{Datta2005,PhysRevLett.68.2512},
\begin{equation}
\label{current3}
\begin{split}
I_{s}&=e\sum_{s'}\int\frac{d\varepsilon}{2\pi}\{T_{ss'}^{e}(\varepsilon)[f_{s}(\varepsilon+eU_{s})-f_{s'}(\varepsilon+eU_{s'})]\\
&+T_{ss'}^{h}(\varepsilon)[f_{s}(\varepsilon+eU_{s})-f_{s'}(\varepsilon-eU_{s'})]\}.\\
\end{split}
\end{equation}
It is worth noting that due to the emergence of MFs, there exists a new transmission channel (hole-channel $T_{ss'}^{h}$ in Eq. (\ref{current3})) in Majorana transport. This leads to the deviation from ordinary tunneling transport in the normal nanowire \cite{Caroli1971,PhysRevLett.68.2512} and renders the left and right currents asymmetric, $I_{L}\ne-I_{R}$. For example, when $\Gamma_{L}=\Gamma_{R}=\Gamma$, the quantity $J=I_{L}+I_{R}$ is shown in Fig. \ref{cas}. We observe that only when $U_{L}=-U_{R}$ ($V=-2U_{L}$, where $V=U_{R}-U_{L}$), $J$ is zero; otherwise, $J\ne0$ in the $\Gamma\sim V$ plane. It is easy to check that when the hole transmission function $T_{ss'}^{h}$ vanishes, the current symmetry is recovered, $I_{L}=-I_{R}$. This scenario happens in the normal semiconducting nanowire \cite{Caroli1971,PhysRevLett.68.2512}, the Anderson model \cite{PhysRev.165.566,Bonca2007} and the quantum dot systems \cite{Gordon1998,Wiel2000}. Therefore, the current asymmetry is a unique feature in the Majorana transport and may be served as an indicator of the emergence of the Majorana fermion. In the Majorana transport, we should define the current going through the wire as $I=\frac{1}{2}(I_{L}-I_{R})$. When the hole transmission function $T_{ss'}^{h}=0$, the definition reduces to the usual one.

\begin{figure}
\begin{tabular}{cc}
\includegraphics[width=3.90cm]{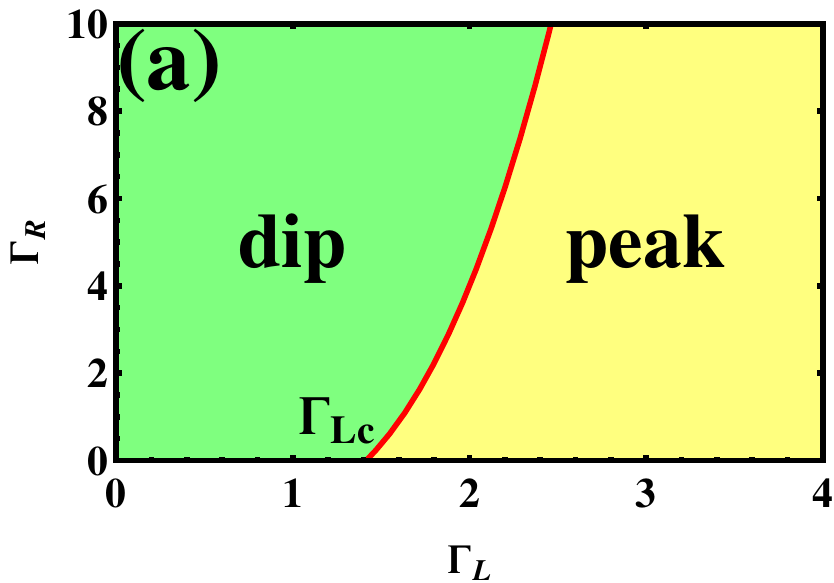} &
\includegraphics[width=4cm]{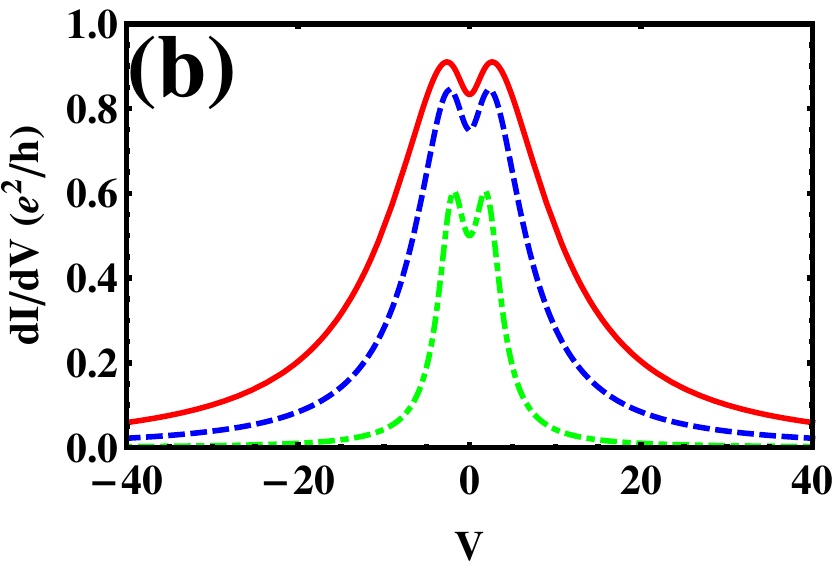} \\
\includegraphics[width=4cm]{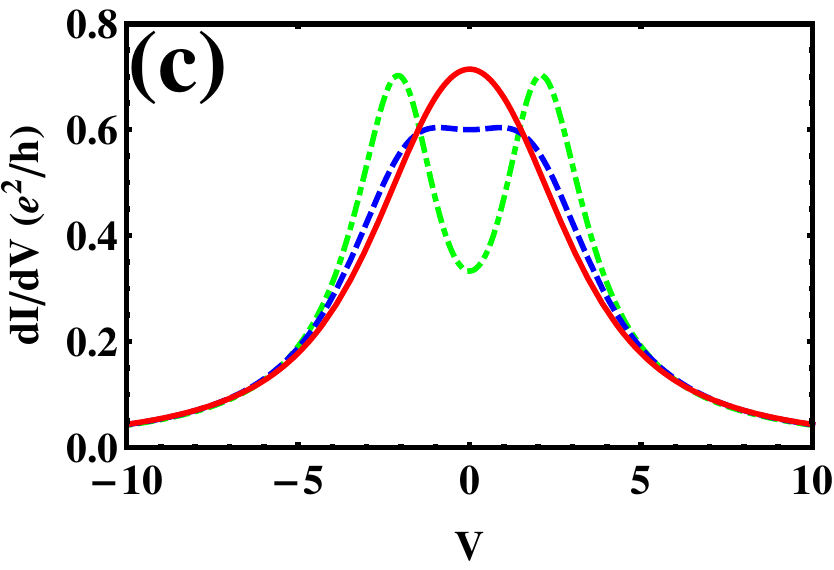} &
\includegraphics[width=4cm]{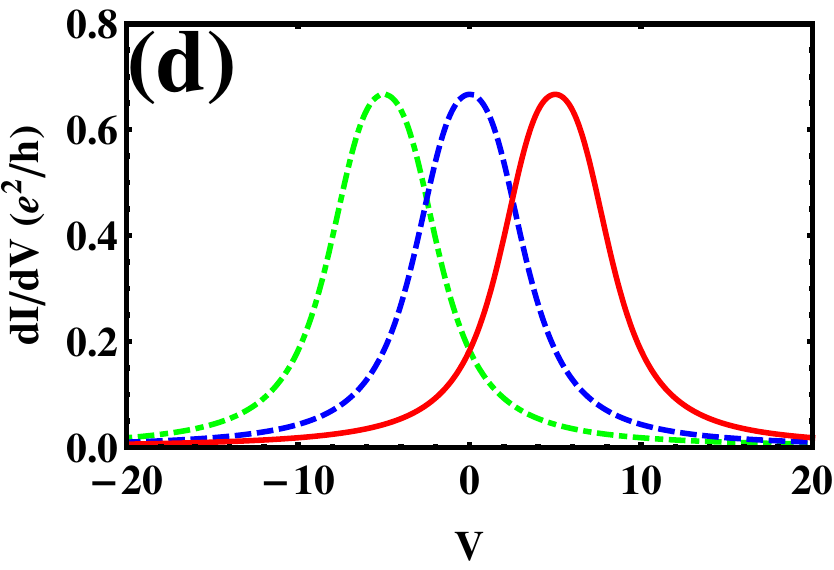} \\
\includegraphics[width=4cm]{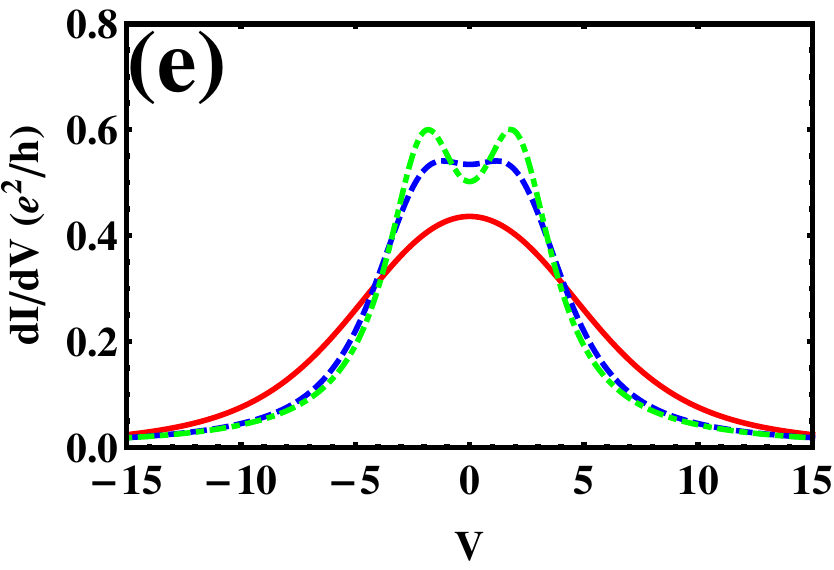} &
\includegraphics[width=4cm]{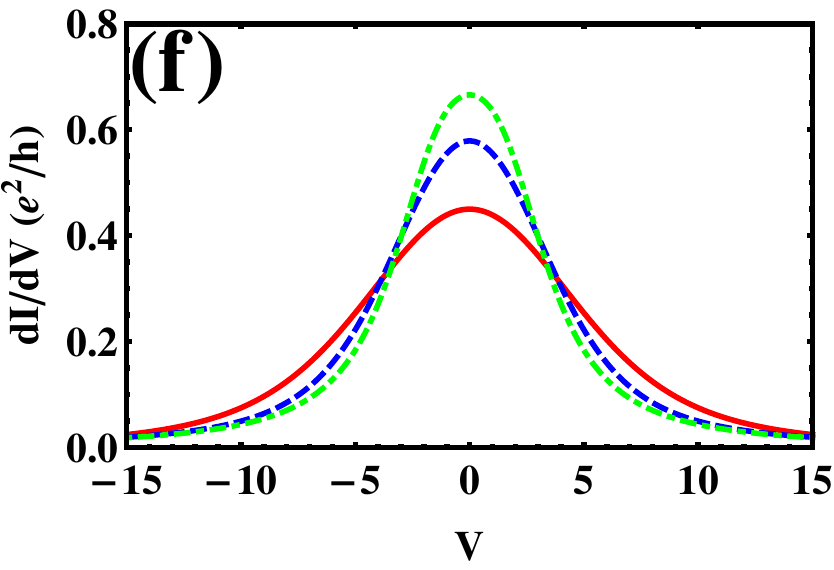} \\
\end{tabular}
\caption{(color online). Conductance for the \textit{dc} voltage.
We set $t=1$ in all figures. (a) shows the critical line for the
dip-peak transition in the $\Gamma_{R}\sim\Gamma_{L}$ plane. The
parameters in (b) are $U_{L}=0$, $\Gamma_{L}=1$ and $\Gamma_{R}=1$ (green
dot-dashed), $3$ (blue dashed), $5$ (red solid); in (c) are $U_{L}=0$, $\Gamma_{R}=1$ and $\Gamma_{L}=0.5$ (green
dot-dashed), $1.5$ (blue dashed), $2.5$ (red solid); in (d) are $\Gamma_{L}=2$, $\Gamma_{R}=1$ and $U_{L}=5$ (green
dot-dashed), $0$ (blue dashed), $-5$ (red solid); in (e) are $U_{L}=0$, $\Gamma_{L}=\Gamma_{R}=1$ and $\beta_{L}=\beta_{R}=10$ (green dot-dashed), $1.5$ (blue dashed), $0.5$ (red solid); in (f) are $U_{L}=0$, $\Gamma_{L}=2$,
$\Gamma_{R}=1$ and $\beta_{L}=\beta_{R}=10$ (green dot-dashed), $1$
(blue dashed), $0.5$ (red solid).}\label{statcurr}
\end{figure}

We now investigate differential conductance of the Majorana nanowire
at zero temperature ($\beta_{L/R}=\infty$). Without loss of
generality, $U_{L}$ is fixed and we calculate the
conductance $dI/dV\sim V$, where $V=U_{R}-U_{L}$.
After some straightforward calculations, we obtain that
\begin{equation}
\label{dccond}
\begin{split}
\frac{dI}{dV}=&\Big\{\frac{4e^2}{h}[\Gamma_{L}\Gamma_{R}(4t^2+4\Gamma_{L}\Gamma_{R})+e^2(V+U_{L})^2\Gamma_{R}^2]\Big\}\Big/\\
&\Big\{[4t^2-e^2(V+U_{L})^2]^2+(4t^2+4\Gamma_{L}\Gamma_{R})^2-(4t^2)^2\\
&+4e^2(V+U_{L})^2(\Gamma_{L}^2+\Gamma_{R}^2)\Big\}.\\
\end{split}
\end{equation}
Notice that when lead L decouples to the Majorana nanowire ($\Gamma_{L}=U_{L}=0$), the conductance reduces to the one in Ref. \cite{PhysRevB.82.180516}. Interestingly, we observe that a critical line $\Gamma_{R}=\Gamma_{L}^3/t^2-2\Gamma_{L}$ separates the zero-bias
conductance peak from zero-bias conductance dip in the
$\Gamma_{R}\sim\Gamma_{L}$ plane as shown in Fig. \ref{statcurr}(a). For $\Gamma_{L}$, there exists a critical value $\Gamma_{Lc}=\sqrt{2}t$, below which the zero-bias conductance always exhibits a dip as shown in Fig. \ref{statcurr}(b). Above this threshold, the zero-bias conductance
undergoes a transition from dip to peak as shown in Fig. \ref{statcurr}(c). We also find that the ZBP becomes larger as the level broadening is increased. The $U_{L}$ dependence of conductance is also shown in Fig. \ref{statcurr}(d). It is easy to see that only when $U_{L}=0$, the peak is zero-bias, otherwise there is a shift in the $V$ direction. We also study the finite temperature effects as depicted in Fig. \ref{statcurr}(e) and \ref{statcurr}(f). As the temperature is increased, the scattering process occurs more frequently, thereby leading to a reduction of the conductance. The competing effect of voltage and temperature can be seen from the intersection of the conductance profiles as well. Notice that even in the dip region of Fig. \ref{statcurr}(a), the dip can become a peak at zero-bias voltage as the temperature is increased as shown in Fig. \ref{statcurr}(e). Although the ZBP above is consistent with the Majorana interpretation, other mechanisms such as impurity, disorder, \cite{PhysRevLett.88.226805,PhysRevB.68.125312,PhysRevB.79.161307} or zero-bias anomaly of Kondo physics \cite{PhysRev.165.566,Bonca2007,Gordon1998,Wiel2000} cannot be completely ruled out. In these cases, the currents at the left and right leads remain symmetric, while in the tunneling transport involving Majorana fermion, the currents are asymmetric. Therefore, the current asymmetry $J$ can be served as an auxiliary criterion for confirming the existence of Majorana fermion in tunneling experiment.

\subsection{\textit{ac} current response}

\begin{figure}
\begin{tabular}{cc}
\includegraphics[width=4cm]{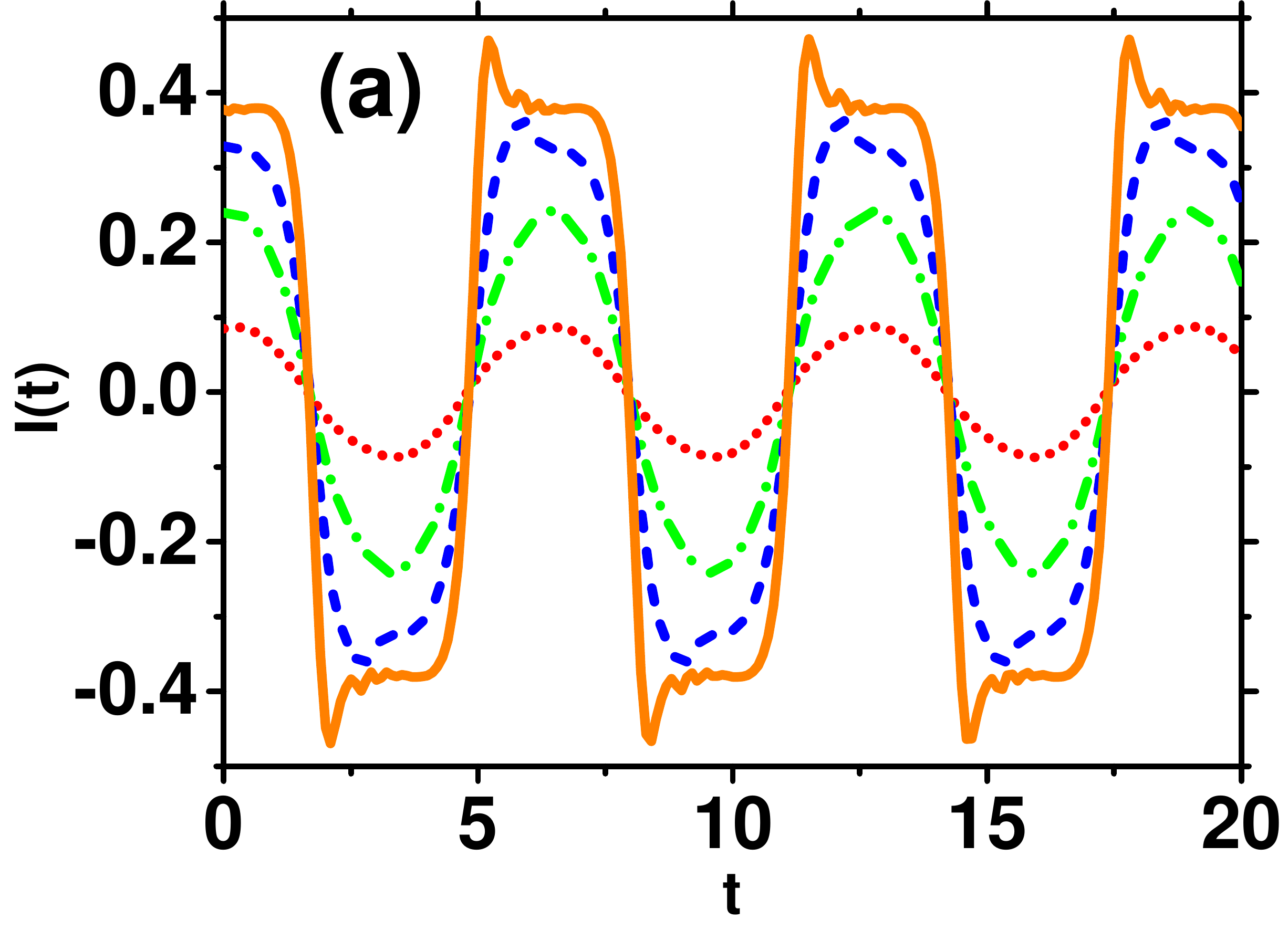} &
\includegraphics[width=4cm]{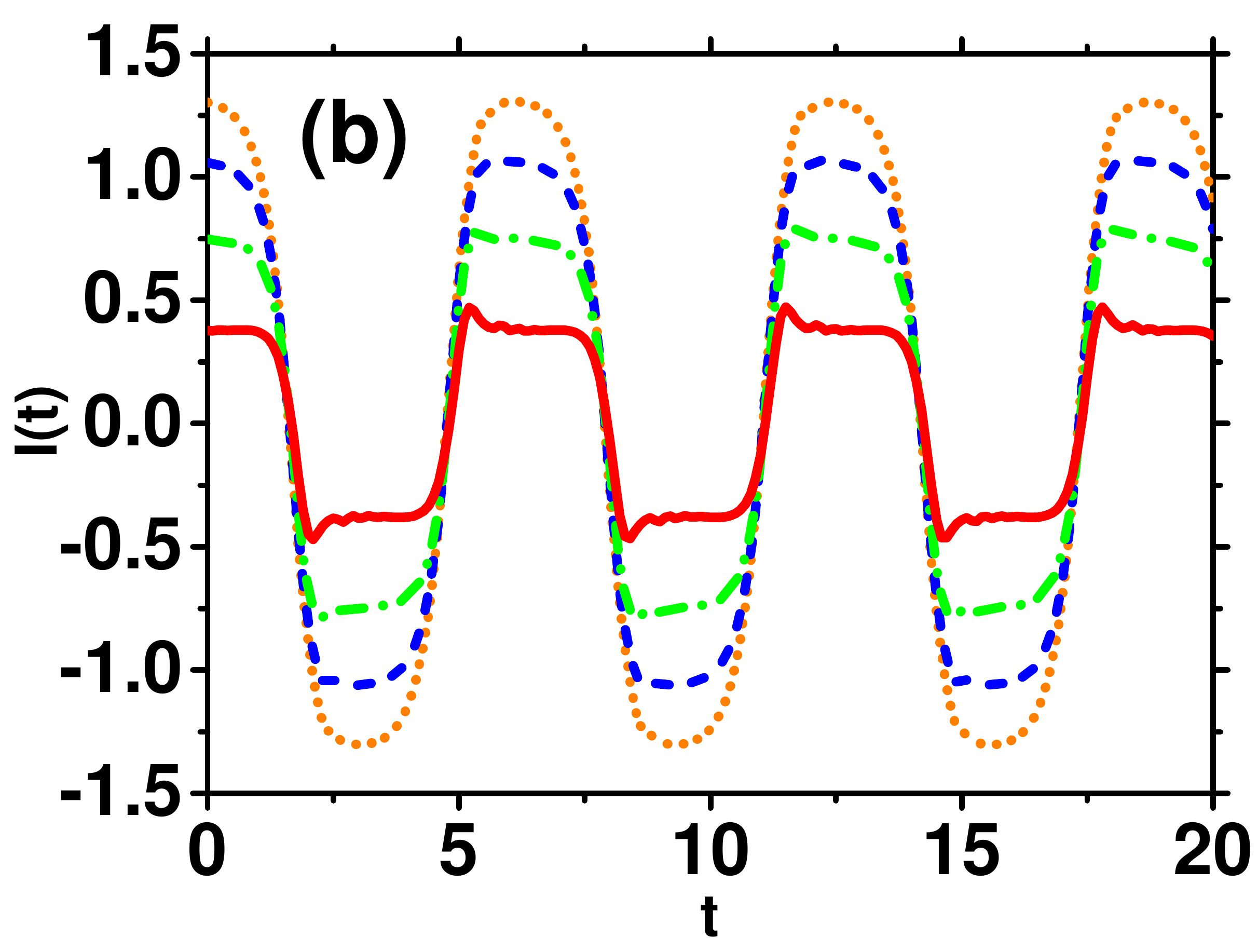} \\
\includegraphics[width=4cm]{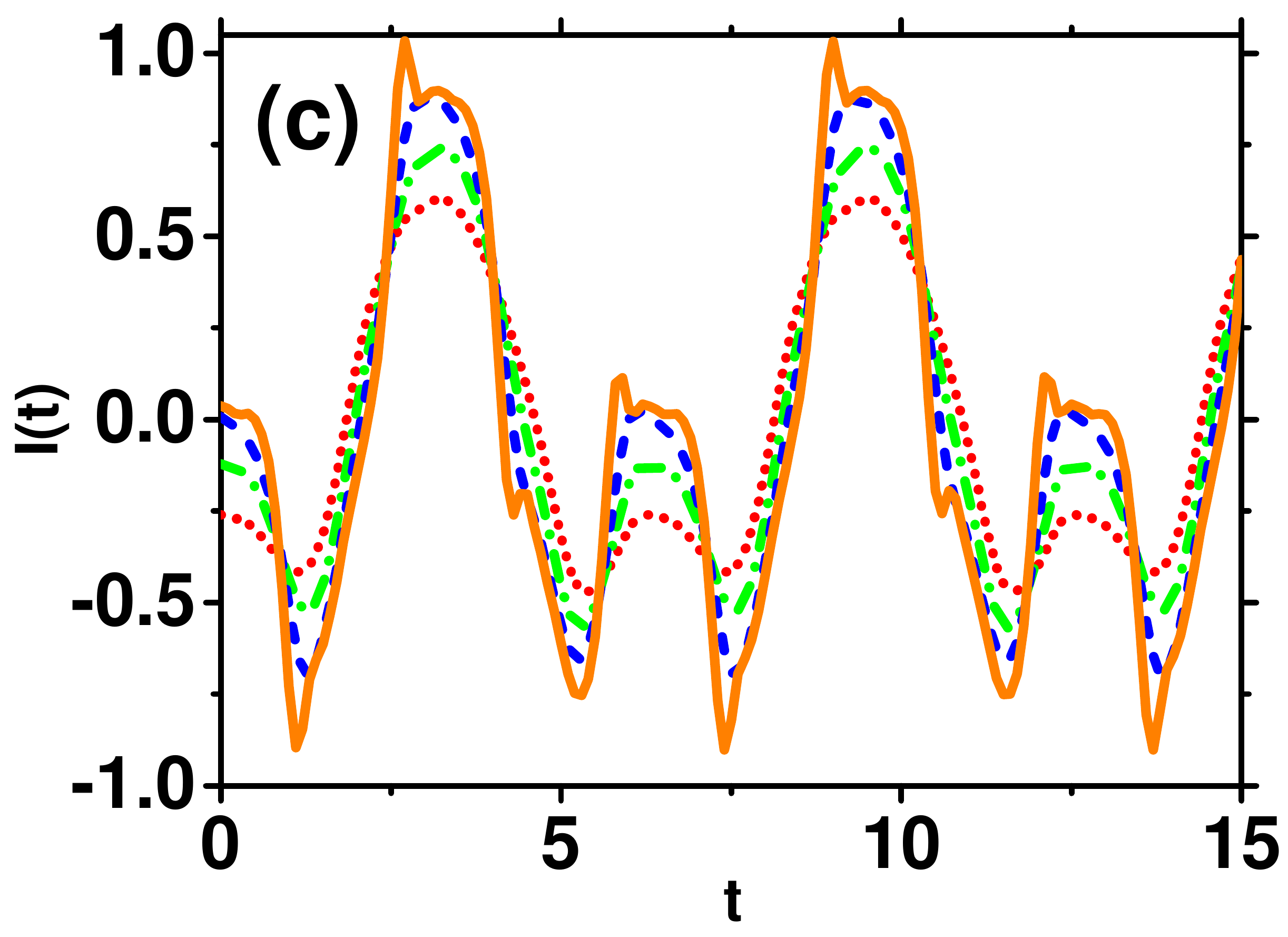} &
\includegraphics[width=4cm]{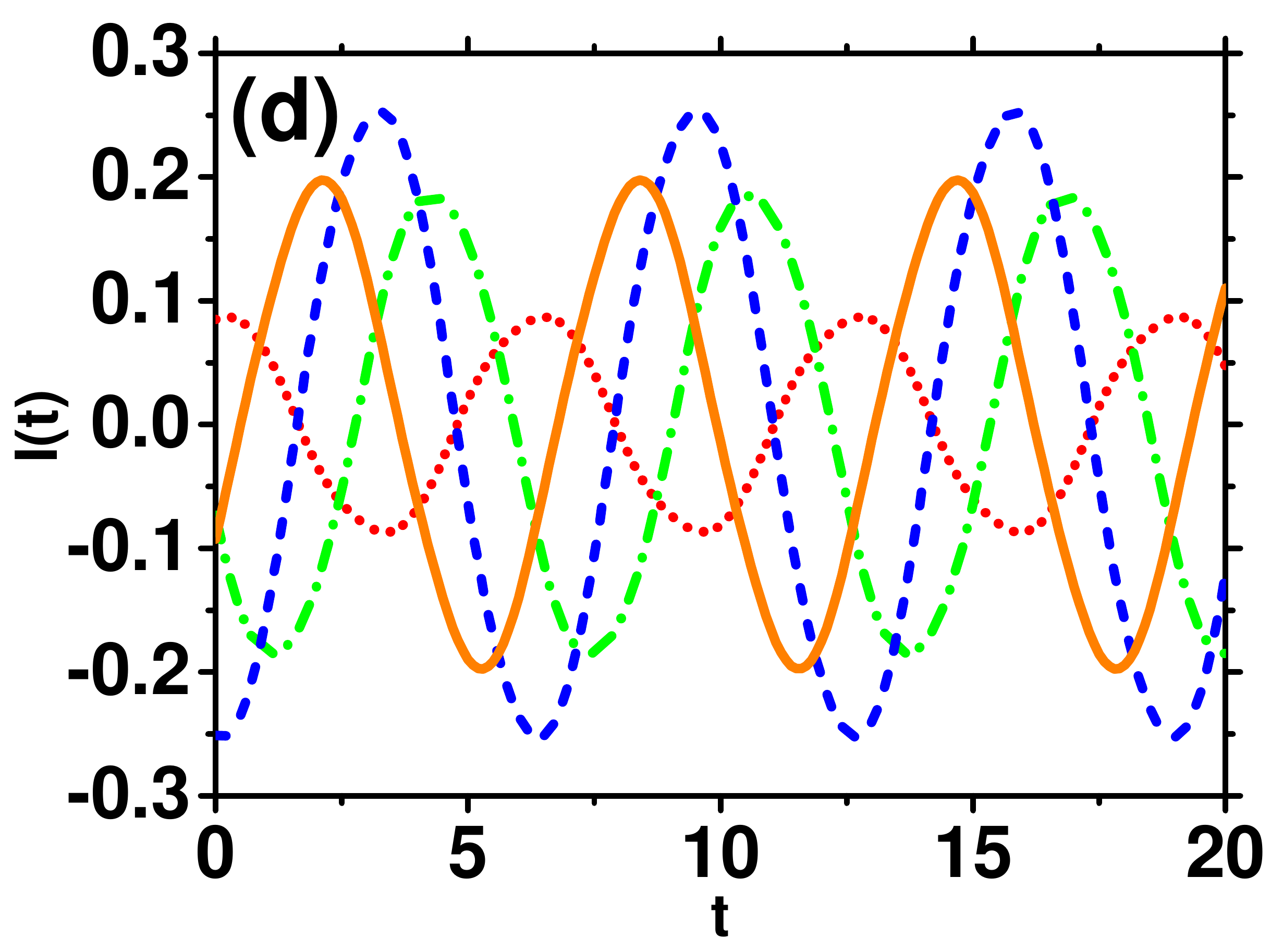} \\
\includegraphics[width=4cm]{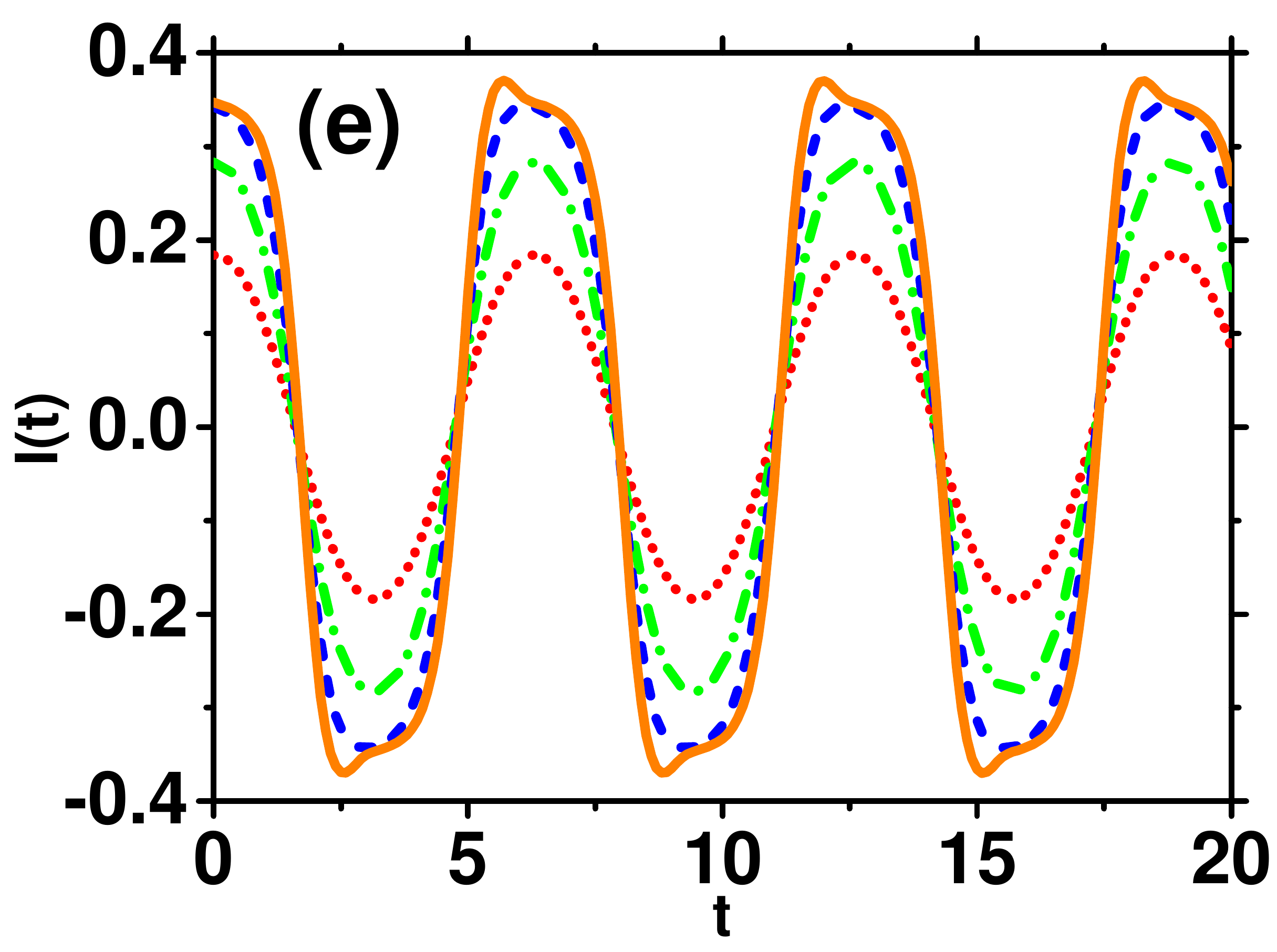} &
\includegraphics[width=4.15cm]{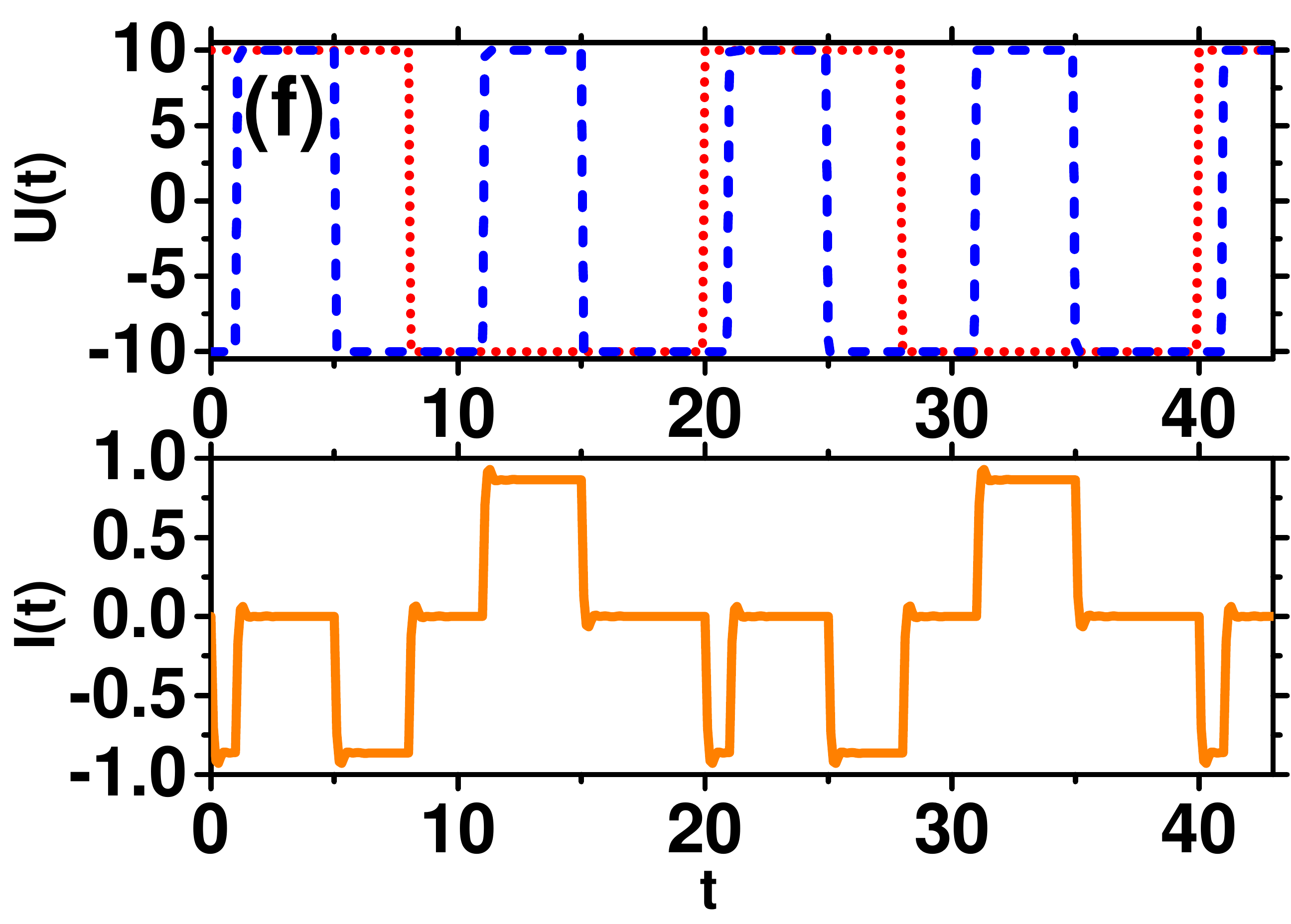} \\
\end{tabular}
\caption{(color online). Current response to the \textit{ac} voltage. (a) response to the change of $U_{R}$ with $\omega_{L}=\omega_{R}=1$ and $U_{R}=2$(dot), $4$ (dash-dot), $8$ (dash), $20$ (solid); (b) response to the change of $\Gamma_{R}$ with $\Gamma_{R}=4$(dot), $3$ (dash), $2$ (dash-dot), $1$ (solid); (c) response to the change of $U_{R}$ with $\omega_{L}=\omega_{R}/2=1$ and $U_{R}=2$(dot), $4$ (dash-dot), $8$ (dash),
$20$ (solid); (d) response to the change of $\phi$ with $\phi=0$(dot), $\pi/2$ (dash-dot), $\pi$ (dash), $3\pi/2$ (solid); (e) response to the change of temperature with $\beta_{L}=\beta_{R}=0.1$(dot), $0.2$ (dash-dot), $0.4$ (dash), $1$ (solid); (f) response to rectangular signals. The dotted line and dashed line in the upper plane are the voltage signals $U_{L}(t)$ and $U_{R}(t)$, respectively. The lower plane is the current response.}\label{dyncurr}
\end{figure}

We turn to consider the current response to the \textit{ac} voltages. The harmonic voltages at the two ends of the nanowire are $U_{L}(t)=U_{L}\cos{\omega_{L}t}$ and $U_{R}(t)=U_{R}\cos{(\omega_{R}t+\phi)}$ respectively. When the voltage $U_{R}$ is enhanced, the current becomes less and less harmonic and finally saturates at high voltage as shown in Fig. \ref{dyncurr}(a). The larger the level broadening $\Gamma_{R}$ is, the stronger the coupling between lead and nanowire is. This leads to a higher current response as shown in Fig. \ref{dyncurr}(b). In Fig. \ref{dyncurr}(c), we study the influence of frequency difference of input signals and find that a more complicated periodic pattern appears. The effect of phase difference is given in Fig. \ref{dyncurr}(d). It is shown that the current response hits the peak when the two voltage signals are out-of-phase. We also study the temperature effect in Fig. \ref{dyncurr}(e) and get similar results as the \textit{dc} case. The response to rectangular \textit{ac} voltages are depicted in Fig. \ref{dyncurr}(f). The upper plane is the voltage signals and the lower plane is the current response. It can be expected that in each plateau, the current response is the same as the \textit{dc} case.

\section{interaction and disorder effects on the Majorana transport}

The interaction and disorder effects on the topological property of the wire are significant. The disorder will destroy the transitional symmetry of the wire. We need to consider the Hamiltonian of the wire Eq. (\ref{H0}) in the real space,
\begin{equation}
\begin{split}
&H_{0}=\int dx\psi^{\dag}[-\frac{\partial_{x}^{2}}{2m}-\mu-i(\alpha\sigma^{y}+\beta\sigma^{x})\partial_{x}+V_{z}\sigma^{z}]\psi,\\
&H_{\Delta}=\Delta\int dx(\psi_{\uparrow}\psi_{\downarrow}+\text{H.c.}).\\
\end{split}
\end{equation}
We will use the bosonization method \cite{Giamarchi,PhysRevB.37.325,PhysRevB.87.060504} to discuss the interaction and disorder effects in the wire. Generally, in one dimension, the localization length due to the disorder is of the order of the mean free path. It means that after bumping a couple of times on the impurities the electrons are localized \cite{Giamarchi} and the wire becomes insulator. However, when the superconducting pairing satisfies the condition Eq. (\ref{pfaffian}), the wire is in the topological superconducting phase.
This competing mechanism can be quantitatively studied by the renormalization analysis of the density-density correlation function of the wire.

\subsection{brief introduction of bosonization}

\subsubsection{left and right movers representation}

To obtain the low-energy properties of the wire, we can deal with excitations close to the Fermi surface. Since the chemical potential $\mu$ lies within the gap, only the lower band $\varepsilon_{-}(k)$ is activated and there are only two Fermi points $\pm k_{F}$ in the energy spectrum. We can linearize the dispersion relation close to each Fermi points. In one dimension, because the low-energy particle-hole excitations have both well-defined momentum and energy, this will lead to two species of fermions: left and right moving fermions. We then replace the original model by one where the energy spectrum is purely linear. This is nothing but assuming that the density of states is constant.

We start with writing the field operator of the Hamiltonian in the left and right movers representation. By the unitary transform $a_{ks}=\sum_{\nu}\langle s|\chi_{\nu}(k)\rangle a_{k\nu}$ where $s=\uparrow,\downarrow$ and $\nu=\pm$, we transform the field operator $\psi_{s}(x)=\frac{1}{\sqrt{2\pi}}\int dk e^{ikx}a_{ks}$ from the spin basis to the band basis and neglect the upper band due to the low-energy approximation. Then we can express the field operator in terms of the left and right movers as
\begin{equation}
\label{lrmovers}
\begin{split}
\psi_{s}(x)=&e^{-ik_{F}x}\langle s|\chi_{-}(-k_{F})\rangle\psi_{L}(x)\\
&+e^{ik_{F}x}\langle s|\chi_{-}(k_{F})\rangle\psi_{R}(x),\\
\end{split}
\end{equation}
where the left and right movers are
\begin{equation}
\label{lrmovers1}
\begin{split}
\psi_{L}(x)&=\frac{1}{\sqrt{2\pi}}\int_{-\infty}^{\infty} dk e^{i(k+k_{F})x} d_{k,L},\\
\psi_{R}(x)&=\frac{1}{\sqrt{2\pi}}\int_{-\infty}^{\infty} dk e^{i(k-k_{F})x} d_{k,R}.\\
\end{split}
\end{equation}

We now express the Hamiltonian in the left and right movers representation.  By linearizing the energy spectrum near the Fermi points and neglecting the upper band operator of the kinetic energy term $H_0$, we have
\begin{equation}
\label{kineticLR}
\begin{split}
H_{0}=i\nu_{F}\int dx[\psi_{L}^{\dag}(x)\partial_{x}\psi_{L}(x)-\psi_{R}^{\dag}(x)\partial_{x}\psi_{R}(x)],\\
\end{split}
\end{equation}
where the Fermi velocity is $\nu_{F}=\frac{k_{F}}{m}-\frac{(\alpha^{2}+\beta^{2})k_{F}}{\sqrt{(\alpha^{2}+\beta^{2})k_{F}^{2}+V_{z}^{2}}}$.
Similarly, substituting Eq. (\ref{lrmovers}) into the $s$-wave superconducting term $H_{\Delta}$, we obtain
\begin{equation}
\label{SCterm}
\begin{split}
H_{\Delta}&=\Delta\sin{\gamma_{k_{F}}}\int dx[\psi_{L}(x)\psi_{R}(x)+\psi_{R}^{\dag}(x)\psi_{L}^{\dag}(x)].\\
\end{split}
\end{equation}
Here we only keep the slowly varying terms and the oscillating terms have been neglected \cite{Giamarchi}.
We next consider the Coulomb interaction which can be
formulated as
\begin{equation}
\begin{split}
H_{\text{int}}=\int dx dx' V(x-x')\rho(x)\rho(x'),\\
\end{split}
\end{equation}
where the electron density operator is $\rho(x)=\sum_{s=\uparrow,\downarrow}\psi_{s}^{\dag}(x)\psi_{s}(x)$. In the momentum space, the interaction can be recast into
\begin{equation}
\begin{split}
H_{\text{int}}=\frac{1}{2\Omega}\sum_{k,s,k',s',q}V(q)a_{k+q,s}^{\dag}a_{k'-q,s'}^{\dag}a_{k',s'}a_{k,s}.\\
\end{split}
\end{equation}
Similarly, we put the Hamiltonian in the band basis $(a_{k+},a_{k-})$ and neglect the upper-band operators, then the interaction term becomes
\begin{equation}
\begin{split}
H_{\text{int}}=\frac{1}{2\Omega}\sum_{k,k',q}&V(q)\cos{\frac{\gamma_{k+q}-\gamma_{k}}{2}}\cos{\frac{\gamma_{k'-q}-\gamma_{k'}}{2}}\\
&d_{k+q}^{\dag}d_{k'-q}^{\dag}d_{k'}d_{k}.\\
\end{split}
\end{equation}
One should remind that the most efficient processes in the interaction are the ones that can act close to the Fermi surface.
Particularly in one dimension, it is worth noting that the Fermi surface is reduced to two points $\pm k_{F}$ thus allows us to decompose the interaction into three scattering processes. The first one is exchange scattering, where two electrons moving in the same direction collide and exchange their velocities; the second one is forward scattering, where two electrons moving in the opposite directions collide and keep moving in their original directions; the third one is the backward scattering, where two electrons moving in the opposite directions collide and move backward. Notice that the wave vector $q$ for the forward and exchange processes is $\sim0$, and $\sim\pm2k_{F}$ for the backward scattering. It is easy to see that the forward and backward scattering processes are identical for the spinless fermion as the particles are indiscernible. Therefore, the interaction Hamiltonian can be expressed as the sum of the above scattering processes:
\begin{equation}
\begin{split}
H_{\text{int}}&=\frac{V(0)}{2\Omega}\sum_{k\in R,k'\in R,q\sim0}d_{k+q}^{\dag}d_{k'-q}^{\dag}d_{k'}d_{k}\\
&+\frac{V(0)}{\Omega}\sum_{k\in R,k'\in L,q\sim0}d_{k+q}^{\dag}d_{k'-q}^{\dag}d_{k'}d_{k}\\
&+(R \leftrightarrow L),\\
\end{split}
\end{equation}
Using the definitions of the left and right movers Eq. (\ref{lrmovers1}), the interaction can be written as
\begin{equation}
\label{intLR}
\begin{split}
H_{\text{int}}&=g_{2}\int dx \psi_{R}^{\dag}(x)\psi_{R}(x)\psi_{L}^{\dag}(x)\psi_{L}(x)\\
&+\frac{g_{4}}{2}\int dx \psi_{R}^{\dag}(x)\psi_{R}(x)\psi_{R}^{\dag}(x)\psi_{R}(x)\\
&+\frac{g_{4}}{2}\int dx \psi_{L}^{\dag}(x)\psi_{L}(x)\psi_{L}^{\dag}(x)\psi_{L}(x),\\
\end{split}
\end{equation}
where $g_{2}/2=g_{4}=V(0)$.
Now we turn to study the disorder term. When the impurities are weak and dense enough so that the effect of each impurity is negligible, they can only act collectively. In this case, there are many impurities in a volume small compared to the scale of variation of the physical quantities but large compared to the distance between impurities. Physically it means that one can replace the original disorder by a coarse grained version. This coarse grained disorder is equivalent to a Gaussian disorder \cite{Giamarchi} due to the central limit theorem. The disorder potential $U(x)$ can be treated as a random chemical potential on the impurity sites. Thus the disorder term can be formulated as
\begin{equation}
\begin{split}
H_{\text{dis}}=\int dx U(x)\rho(x),\\
\end{split}
\end{equation}
where the disorder potential satisfies Gaussian distribution
\begin{equation}
\begin{split}
p(U)=\exp{\big[-\frac{1}{2D}\int dx U^{2}(x)\big]}.\\
\end{split}
\end{equation}
Here we assume that the impurity potential is short-range so that $\langle U(x)U(x')\rangle=D\delta(x-x')$. Fourier transforming the Hamiltonian and projecting it onto the lower band, we have
\begin{equation}
\begin{split}
H_{\text{dis}}=\frac{1}{\Omega}\sum_{k,q}U(q)\cos{\frac{\gamma_{k+q}-\gamma_{k}}{2}}d_{k+q}^{\dag}d_{k}.\\
\end{split}
\end{equation}
Again, the most important processes are the ones close to the Fermi surface. In one dimension, the disorder term can be thus approximated as
\begin{equation}
\begin{split}
H_{\text{dis}}&=\frac{1}{\Omega}\sum_{q\sim0}U(q)\sum_{k\sim\pm k_{F}}d_{k+q}^{\dag}d_{k}\\
&+\frac{\cos{\gamma_{k_{F}}}}{\Omega}\sum_{q\sim2k_{F}}U(q)\sum_{k\sim-k_{F}}d_{k+q}^{\dag}d_{k}\\
&+\frac{\cos{\gamma_{k_{F}}}}{\Omega}\sum_{q\sim-2k_{F}}U(q)\sum_{k\sim k_{F}}d_{k+q}^{\dag}d_{k}.\\
\end{split}
\end{equation}
In the left and right movers representation, we have
\begin{equation}
\label{disorderLR}
\begin{split}
H_{\text{dis}}&=\int dx\eta(x)[\psi_{R}^{\dag}(x)\psi_{R}(x)+\psi_{L}^{\dag}(x)\psi_{L}(x)]\\
&+\cos{\gamma_{k_{F}}}\int dx[\xi(x)\psi_{L}^{\dag}(x)\psi_{R}(x)+\xi^{*}(x)\psi_{R}^{\dag}(x)\psi_{L}(x)],\\
\end{split}
\end{equation}
where $\eta(x)=\frac{1}{\Omega}\sum_{q\sim0}U(q)e^{iqx}$ and
$\xi(x)=\frac{1}{\Omega}\sum_{q\sim0}U(q-2k_{F})e^{iqx}$ are two independent Gaussian random variables. Note that $\eta(x)$ is real and $\xi(x)$ is complex. The correlation relations are $\langle\eta(x)\eta(x')\rangle=\langle\xi(x)\xi^{*}(x')\rangle=D\delta(x-x')$, and zero otherwise.

\subsubsection{bosonization of the Majorana nanowire}

The Abelian bosonization formula \cite{Giamarchi,PhysRevB.87.060504,PhysRevLett.109.146403,PhysRevB.78.054436} is given by
\begin{equation}
\label{bosonizationformula}
\begin{split}
\psi_{L}(x)&=\frac{1}{\sqrt{2\pi\alpha}}e^{-i\sqrt{4\pi}\phi_{L}(x)},\\
\psi_{R}(x)&=\frac{1}{\sqrt{2\pi\alpha}}e^{i\sqrt{4\pi}\phi_{R}(x)},\\
\end{split}
\end{equation}
where $\psi_{L/R}(x)$ is the massless Dirac
field (fermionic) as shown in Eq. (\ref{lrmovers1}), and $\phi_{L/R}(x)$ is massless Klein-Gordon field (bosonic). $\alpha$ is the short-range cutoff for the convergence of the continuum theory.

Using the formula Eq. (\ref{bosonizationformula}), the kinetic energy Eq. (\ref{kineticLR}) can be bosonized as
\begin{equation}
\begin{split}
H_{0}&=\nu_{F}\int dx \{[\partial_{x}\phi_{L}(x)]^{2}+[\partial_{x}\phi_{R}(x)]^{2}\}.\\
\end{split}
\end{equation}
We define two new variables,
$\phi_{L}=\frac{-1}{\sqrt{4\pi}}(\theta+\varphi)$ and
$\phi_{R}=\frac{1}{\sqrt{4\pi}}(\theta-\varphi)$, where the commutation relation is $[\theta(x),\varphi(y)]=i\pi\text{sgn}(y-x)/2$. The Hamiltonian $H_0$ then becomes
\begin{equation}
\begin{split}
H_{0}=\frac{\nu_{F}}{2\pi}\int
dx[(\partial_{x}\theta)^{2}+(\partial_{x}\varphi)^{2}].
\end{split}
\end{equation}
For the $s$-wave superconducting term, substituting Eq. (\ref{bosonizationformula}) into Eq. (\ref{SCterm}), after bosonization it can be written as
\begin{equation}
\begin{split}
H_{\Delta}=\frac{\Delta\sin{\gamma_{k_{F}}}}{\pi\alpha}\int dx \cos{2\theta}.\\
\end{split}
\end{equation}
Similarly, the Coulomb interaction Eq. (\ref{intLR}) in terms of the field $\theta$ and $\varphi$ is
\begin{equation}
\begin{split}
H_{\text{int}}&=\frac{g_{2}}{4\pi^2}\int dx [(\partial_{x}\varphi)^{2}-(\partial_{x}\theta)^{2}]\\
&+\frac{g_{4}}{4\pi^2}\int dx [(\partial_{x}\varphi)^{2}+(\partial_{x}\theta)^{2}].\\
\end{split}
\end{equation}
The bosonic form of the disorder term Eq. (\ref{disorderLR}) is given by
\begin{equation}
\label{disorderboson}
\begin{split}
H_{\text{dis}}&=-\frac{1}{\pi}\int dx [\eta(x) \partial_{x}\varphi]\\
&+\frac{\cos{\gamma_{k_{F}}}}{2\pi\alpha}\int dx [\xi(x)e^{-i2\varphi}+\xi^{*}(x)e^{i2\varphi}].\\
\end{split}
\end{equation}
Notice that in one dimension, the effect of Coulomb interaction just leads to the reparameterization of the kinetic energy $H_0$. The interaction can be absorbed into the kinetic energy, then we arrive at the following Hamiltonian for the Luttinger liquid,
\begin{equation}
\label{Luttingerliquid}
\begin{split}
H_{\text{Lutt}}=H_{0}+H_{\text{int}}=\frac{1}{2\pi}\int dx\left[uK(\partial_{x}\theta)^{2}+\frac{u}{K}(\partial_{x}\varphi)^{2}\right],\\
\end{split}
\end{equation}
where the Luttinger parameters are
\begin{equation}
\begin{split}
&u=\nu_{F}\left[(1+y_{4})^{2}-y_{2}^{2}\right]^{1/2},\\
&K=\left(\frac{1+y_{4}-y_{2}}{1+y_{4}+y_{2}}\right)^{1/2}.\\
\end{split}
\end{equation}
Here $y_{2}=\frac{g_{2}}{2\pi\nu_{F}}$ and
$y_{4}=\frac{g_{4}}{2\pi\nu_{F}}$. Furthermore, we observe that the first term in Eq. (\ref{disorderboson}) is equivalent to a random gauge potential which can also be absorbed into the Hamiltonian of Luttinger liquid Eq. (\ref{Luttingerliquid}) via replacing $\varphi(x)$ by $\varphi(x)-\frac{K}{u}\int^{x}dy \eta(y)$. Finally, we achieve the bosonized Hamiltonian of the interacting wire with Gaussian disorder $H_{\text{nw}}=H_{\text{sG}}+H_{\text{bws}}$, where
\begin{equation}
\begin{split}
H_{\text{sG}}&=\frac{1}{2\pi}\int dx\left[uK(\partial_{x}\theta)^{2}+\frac{u}{K}(\partial_{x}\varphi)^{2}\right]\\
&+\frac{\Delta\sin{\gamma_{k_{F}}}}{\pi\alpha}\int dx \cos{2\theta},\\
H_{\text{bws}}&=\frac{\cos{\gamma_{k_{F}}}}{2\pi\alpha}\int dx [\xi(x)e^{-i2\varphi}+\xi^{*}(x)e^{i2\varphi}].\\
\end{split}
\end{equation}
The Hamiltonian $H_{\text{sG}}$ is the sine-Gordon Hamiltonian which is well-known to have Kosterlitz-Thouless phase transition \cite{Kosterlitz1974}. The superconducting term $\Delta$ couples to the field $\theta(x)$ favoring a superconducting ground state, however, the disorder term couples to the field $\varphi(x)$ and tends to pin the charge density to the disorder potential \cite{PhysRevLett.109.146403}. Therefore, we can expect the disorder system undergoes a topological phase transition as the interplay of superconductivity and disorder.

\subsection{influence on the Majorana transport}

\begin{figure}
\begin{tabular}{cc}
\includegraphics[width=4cm]{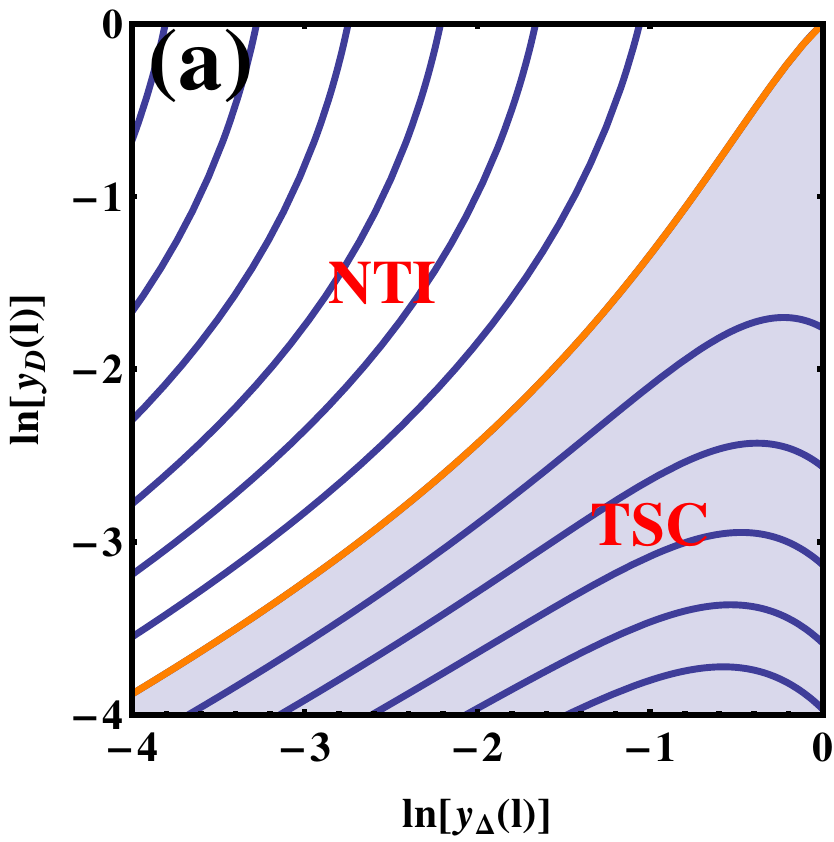} &
\includegraphics[width=4cm]{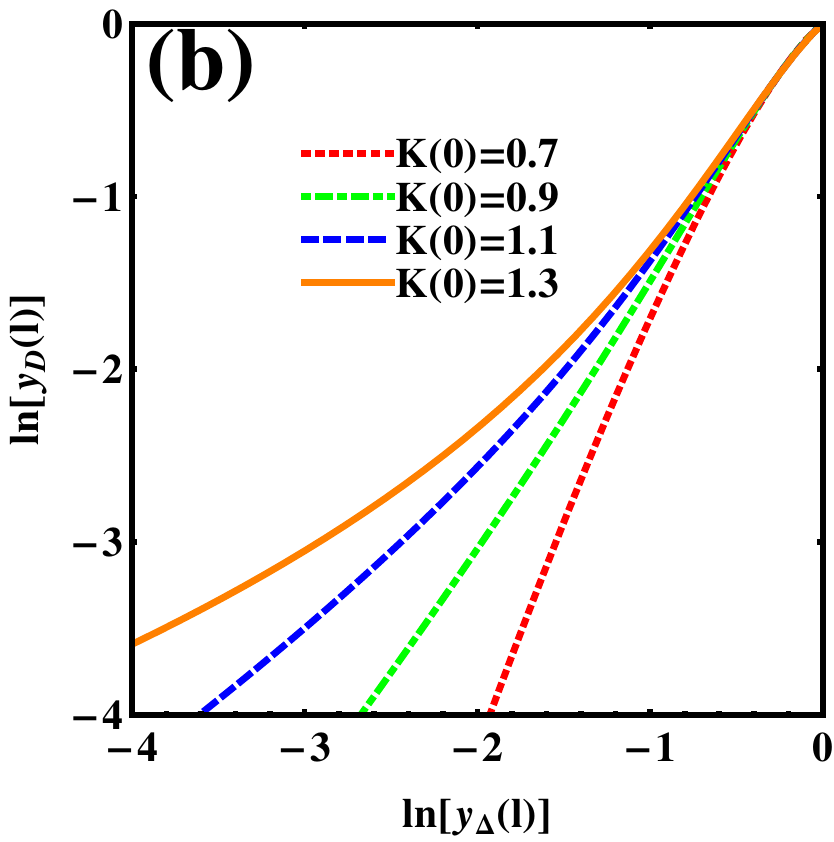} \\
\end{tabular}
\caption{(color online). (a) RG flows of $y_{D}(l)$ and $y_{\Delta}(l)$, where $K(0)=1.2$. NTI and TSC are short for non-topological insulator and topological superconductor. The orange line is the phase boundary. (b) phase boundaries with different $K(0)$s.}\label{RGflow}
\end{figure}

In general, we can use the perturbation theory to calculate the correlation function, for instance, $R(r_{1}-r_{2})=\langle e^{ia\sqrt{2}\varphi(r_{1})}e^{-ia\sqrt{2}\varphi(r_{2})}\rangle_{H_{\text{nw}}}$, to study the physical property of the interacting disorder wire. The average for a system with a disorder potential $U$ can be treated by the replica method \cite{Giamarchi} as follows,
\begin{equation}
\label{aveO}
\begin{split}
\langle O&(\varphi)\rangle_{H_{\text{nw}}}=\\
&\lim_{n\rightarrow0}\frac{\int\mathcal{D}[\xi,\xi^{*}]p(\xi,\xi^{*})\int\prod\limits_{i=1}^{n}\mathcal{D}\varphi_{i}O(\varphi_{1})e^{-\sum\limits_{a=1}^{n}S_{U}(\theta_{a},\varphi_{a})}}{\int\mathcal{D}[\xi,\xi^{*}]p(\xi,\xi^{*})},\\
\end{split}
\end{equation}
where $O(\varphi)$ is some observable of $\varphi$. We need to perform the functional integral over the $n$-copies of replicas $\varphi_{i}$ and the Gaussian distributed random variables $\xi$ and $\xi^{*}$. The Gaussian distributed disorder potential is $p(\xi,\xi^{*})=e^{-\frac{1}{D}\int dx \xi(x)\xi^{*}(x)}$ and the action of the disorder system can be achieved by the Legendre transformation of the Hamiltonian $H_{\text{nw}}$,
\begin{equation}
\begin{split}
S_{U}(\theta_{a},\varphi_{a})=-\int_{0}^{\beta}d\tau\int dx\frac{i}{\pi}\partial_{x}\theta\partial_{\tau}\varphi+\int_{0}^{\beta}d\tau H_{\text{nw}}.\\
\end{split}
\end{equation}
By integrating out the Gaussian random variables $\xi$ and $\xi^{*}$, we arrive at
\begin{equation}
\label{OHnw}
\begin{split}
\langle O(\varphi)\rangle_{H_{\text{nw}}}&=\lim_{n\rightarrow0}\int\prod_{i=1}^{n}\mathcal{D}\varphi_{i}O(\varphi_{1})e^{-S_{\text{eff}}},\\
\end{split}
\end{equation}
where
\begin{equation}
\begin{split}
&S_{\text{eff}}=\sum_{a=1}^{n}\left[S_{0}(\varphi_{a})+S_{\Delta}(\varphi_{a})\right]-\\
&\frac{D\cos^2{\gamma_{k_{F}}}}{(2\pi\alpha)^{2}}\sum_{a,b=1}^{n}\int dxd\tau d\tau'\cos{[2\varphi_{a}(x,\tau)-2\varphi_{b}(x,\tau')]}.\\
\end{split}
\end{equation}
Here $S_{0}=-\int_{0}^{\beta}d\tau\int dx\frac{i}{\pi}\partial_{x}\theta\partial_{\tau}\varphi+\int_{0}^{\beta}d\tau H_{\text{Lutt}}$ and $S_{\Delta}=\int_{0}^{\beta}d\tau H_{\Delta}$. The details of calculation of correlation $R(r_{1}-r_{2})$ by the replica method Eq. (\ref{OHnw}) is given in Appendix \ref{RGC}. After some calculations, we find that the perturbation result is plagued by divergence which is notorious in one dimension. However, although the correlation is infinite, it should be independent of the change of short-range cutoff $\alpha$ because it characterizes the physical properties of the system. This peculiar property suggests the use of the renormalization group method \cite{Giamarchi,PhysRevLett.109.146403}. Particularly, by expanding the superconducting and disorder actions to the first leading order, changing the short-range cutoff $\alpha\rightarrow\alpha e^{l}$ and keeping the correlation function unchanged (see Appendix \ref{RGC} for details), we obtain the RG flows as follows,
\begin{equation}
\begin{split}
&dK/dl=y_{\Delta}^{2}-y_{D}K^{2},\\
&dy_{\Delta}/dl=(2-K^{-1})y_{\Delta},\\ &dy_{D}/dl=2Ky_{D}^{2}-(2K-3+2K^{-1}y_{\Delta}^{2})y_{D},\\
&du/dl=-y_{D}Ku,\\
\end{split}
\end{equation}
where $y_{\Delta}=\alpha\Delta\sin{\gamma_{k_{F}}}/u$ and
$y_{D}=\alpha D\cos^{2}{\gamma_{k_{F}}}/\pi u^{2}$. From the flows of $y_{\Delta}(l)$ and $y_{D}(l)$, we can see that when $K(l)<1/2$, $y_{D}(l)$ is relevant, the system is in the random-pinned change density wave phase; when $K(l)>3/2$, $y_{\Delta}(l)$ is relevant, the system is in the superconducting phase. When $1/2<K(l)<3/2$, both $y_{D}(l)$ and $y_{\Delta}(l)$ are relevant. In order to be consistent with the perturbation condition, the flows can be chosen to stop at $l^{*}$ when $\max{[y_{D}(l^{*}),y_{\Delta}(l^{*})]}=1$. Using this criterion, the phase diagram in this $K(l)$ interval is obtained as shown in Fig. \ref{RGflow}(a). Only when the parameters are in the shadow region, the Majorana fermions remain. In Fig. \ref{RGflow}(b), we plot the phase boundaries with respect to different initial $K(0)$ and find that the topological superconducting phase becomes larger as $K(0)$ increases. Therefore, when $K(l)<1/2$ or when $1/2<K(l)<3/2$ as well as the parameters are in the shadow region of Fig. \ref{RGflow}(a), $y_{\Delta}$ is relevant and the Majorana transport is preserved, otherwise the disorder strength $D$ will destroy the transport.

\section{Summary}

We have used the Keldysh formalism to comprehensively study the multi-lead tunneling in Majorana nanowire with or without short-range Coulomb interaction and disorder. A zero-bias \textit{dc} conductance peak appears in our layout which implies the existence of Majorana fermion and is consistent with previous experiments on InSb nanowire \cite{Mourik2012}. We find that since the MF is a fermion that is its own antiparticle, there exists a hole transmission channel which makes the currents asymmetric at the left and right leads. This current asymmetry may be used as a criterion for detecting the Majorana fermion. For the \textit{ac} voltage, we find that the current response is enhanced in step with the increase of level broadening and the decrease of temperature, and finally saturates at high voltage. The effects of short-range Coulomb interaction and disorder to the Majorana transport have been considered via bosonization method and renormalization group analysis. We find that there is a topological phase transition in the interplay of superconductivity and disorder. In the topological superconducting phase, the Majorana transport remains, otherwise the transport will destroy in the non-topological insulator phase.

\begin{acknowledgments}
This work is partly supported by National Research Foundation and
Ministry of Education, Singapore (Grant No. WBS: R-710-000-008-271). X. Q. Shao is supported by Fundamental Research Funds for the Central Universities under Grant No. 12SSXM001, National Natural Science Foundation of China under Grant No. 11204028, and  the Government of China through CSC.
\end{acknowledgments}

\appendix

\section{Green function and self-energy for Majorana nanowire}
\label{wireGF}

Here we use the equation of motion method to study the Green function of the wire. The Keldysh Green function for the nanowire is defined as $G_{ij}(t,t')=-i\langle
T_{K}\gamma_{i}(t)\gamma_{j}(t')\rangle$. We first consider the time evolution of this Green function. By definition
\begin{equation}
\begin{split}
G_{ij}(t,t')&=-i\theta(t-t')\langle\gamma_{i}(t)\gamma_{j}(t')\rangle\\
&+i\theta(t'-t)\langle\gamma_{j}(t')\gamma_{i}(t)\rangle,\\
\end{split}
\end{equation}
where $\theta(t-t')$ is defined on the Keldysh contour. Then we have
\begin{equation}
\label{evlG}
\begin{split}
\partial_{t}G_{ij}(t,t')=-2i\delta(t-t')\delta_{i,j}-i\langle T_{K}\partial_{t}\gamma_{i}(t)\gamma_{j}(t')\rangle.\\
\end{split}
\end{equation}
Using the Heisenberg equation $i\partial_{t}\gamma_{k}(t)=[\gamma_{k}(t),H]$, where $H$ is the Hamiltonian of the system, it is easy to check that the time evolution of Majorana operator is
\begin{equation}
\label{evlgamma}
\begin{split}
i\partial_{t}\gamma_{k}(t)&=-2i\sum_{i}t_{ik}\gamma_{i}(t)\\
&-\sum_{p,s=L,R}2(V_{pk,s}^{*}c_{p,s}^{\dag}(t)-V_{pk,s}c_{p,s}(t)).\\
\end{split}
\end{equation}
Substituting Eq. (\ref{evlgamma}) into Eq. (\ref{evlG}), after some calculations we have
\begin{equation}
\label{evlG2}
\begin{split}
i\partial_{t}G_{ij}&(t,t')=2\delta(t-t')\delta_{i,j}+2i\sum_{k}t_{ik}G_{kj}(t,t')\\
&+2\sum_{p,s=L,R}(V_{pi,s}G_{pj,s}(t,t')-V_{pi,s}^{*}\overline{G}_{pj,s}(t,t')),\\
\end{split}
\end{equation}
where the Green function for the tunnelings are given by
\begin{equation}
\begin{split}
G_{pj,s}(t,t')&=-i\langle T_{K}c_{p,s}(t)\gamma_{j}(t')\rangle,\\
\overline{G}_{pj,s}(t,t')&=-i\langle
T_{K}c_{p,s}^{\dag}(t)\gamma_{j}(t')\rangle.\\
\end{split}
\end{equation}

Similarly, we can use the EOM method to express these two tunneling Green functions in terms of the wire Green function. for example, we have $(i\partial_{t}-\xi_{p,s})G_{pj,s}(t,t')=\sum_{i}V_{pi,s}^{*}G_{ij}(t,t')$ which leads to the following closed form for $G_{pj,s}(t,t')$,
\begin{equation}
\label{cldG}
\begin{split}
G_{pj,s}(t,t')&=\sum_{i}\int dt''G_{p}^{0}(t,t'')V_{pi,s}^{*}G_{ij}(t'',t'),\\
\end{split}
\end{equation}
and similarly,
\begin{equation}
\label{cldGbar}
\begin{split}
\overline{G}_{pj,s}(t,t')&=-\sum_{i}\int dt''\overline{G}_{p}^{0}(t,t'')V_{pi,s}G_{ij}(t'',t'),\\
\end{split}
\end{equation}
where the free lead Green function is defined as
\begin{equation}
\begin{split}
G_{p}^{0}(t,t')&=-i\langle T_{K}c_{p}(t)c_{p}^{\dag}(t')\rangle_{0},\\
\overline{G}_{p}^{0}(t,t')&=-i\langle T_{K}c_{p}^{\dag}(t)c_{p}(t')\rangle_{0}.\\
\end{split}
\end{equation}
It is easy to check that the free lead Green functions satisfy the point charge source equations,
\begin{equation}
\label{G0}
\begin{split}
(i\partial_{t}-\xi_{p,L})G_{p}^{0}(t,t')&=\delta(t-t'),\\
(i\partial_{t}+\xi_{p,L})\overline{G}_{p}^{0}(t,t')&=\delta(t-t'),\\
\end{split}
\end{equation}
which lead to the closed forms for $G_{pj,s}(t,t')$ and $\overline{G}_{pj,s}(t,t')$ respectively.

Therefore, substituting Eq. (\ref{cldG}) and Eq. (\ref{cldGbar}) into Eq. (\ref{evlG2}), we have
\begin{equation}
\label{evlG3}
\begin{split}
i\partial_{t}G_{ij}(t,t')&=2\delta(t-t')\delta_{i,j}+2i\sum_{k}t_{ik}G_{kj}(t,t')\\
&+2\sum_{k,s=L,R}\int dt''\Sigma_{ik,s}(t,t'')G_{kj}(t'',t'),\\
\end{split}
\end{equation}
where the self-energy is $\mathbf{\Sigma}_{s}=\mathbf{\Sigma}_{s}^{e}+\mathbf{\Sigma}_{s}^{h}$. The electron and hole self-energy are given by
\begin{equation}
\label{KeldyshSE}
\begin{split}
\Sigma_{ik,s}^{e}(t,t'')&=\sum_{p}V_{pi,s}(t)G_{p,s}^{0}(t,t'')V_{pk,s}^{*}(t''),\\
\Sigma_{ik,s}^{h}(t,t'')&=\sum_{p}V_{pk,s}(t)\overline{G}_{p,s}^{0}(t,t'')V_{pi,s}^{*}(t'').\\
\end{split}
\end{equation}

We now study the retarded component of the self-energy $\Sigma_{ik}^{R}=\Sigma_{ik,L}^{eR}+\Sigma_{ik,R}^{eR}+\Sigma_{ik,L}^{hR}+\Sigma_{ik,R}^{hR}$, where
\begin{equation}
\label{retSE}
\begin{split}
\Sigma_{ik,s}^{eR}(t,t'')&=\sum_{p}V_{pi,s}(t)G_{p,s}^{0R}(t,t'')V_{pk,s}^{*}(t''),\\
\Sigma_{ik,s}^{hR}(t,t'')&=\sum_{p}V_{pi,s}^{*}(t)\overline{G}_{p,s}^{0R}(t,t'')V_{pk,s}(t'').\\
\end{split}
\end{equation}
By the wide-band approximation, the retarded self-energy is
\begin{equation}
\label{retSE1}
\begin{split}
\Sigma_{ik,s}^{eR}(t,t'')&=\int\frac{d\varepsilon}{2\pi}[\Gamma_{s}]_{ik}G_{p,s}^{0R}(t,t'').\\
\end{split}
\end{equation}
Substituting the free retarded Green function into Eq. (\ref{retSE1}), the retarded self-energy for electron becomes
\begin{equation}
\begin{split}
\mathbf{\Sigma}_{s}^{eR}(t,t'')&=-\frac{i}{2}\mathbf{\Gamma}_{s}(t)\delta(t-t'').\\
\end{split}
\end{equation}
Similarly, the retarded self-energy for hole is
\begin{equation}
\begin{split}
\mathbf{\Sigma}_{s}^{hR}(t,t'')&=-\frac{i}{2}\mathbf{\Gamma}_{s}^{*}(t)\delta(t-t'').\\
\end{split}
\end{equation}
Therefore, the retarded self-energy for the Majorana nanowire is
\begin{equation}
\label{SER}
\begin{split}
\mathbf{\Sigma}^{R}(t,t'')&=-\frac{i}{2}\mathbf{\Gamma}(t)\delta(t-t''),\\
\end{split}
\end{equation}
where $\mathbf{\Gamma}(t)=\mathbf{\Gamma}_{L}(t)+\mathbf{\Gamma}_{R}(t)+\mathbf{\Gamma}_{L}^{*}(t)+\mathbf{\Gamma}_{R}^{*}(t)$.
Finally, by the analytical continuation \cite{Mahan1990} of Eq. (\ref{evlG3}), we have
\begin{equation}
\begin{split}
i\partial_{t}G_{ij}^{R}(t,t')&=2\delta(t-t')\delta_{ij}+2i\sum_{k}t_{ik}G_{kj}^{R}(t,t')\\
&+2\sum_{k}\int dt''\Sigma_{ik}^{R}(t,t'')G_{kj}^{R}(t'',t').\\
\end{split}
\end{equation}
Substituting Eq. (\ref{SER}) into it, we arrive at $[i\partial_{t}-2i\mathbf{t}+i\mathbf{\Gamma}(t)]\mathbf{G}^{R}(t,t')=2\delta(t-t')$, which leads to the solution to the retarded Green function
\begin{equation}
\begin{split}
\mathbf{G}^{R}(t,t')&=-2i\theta(t-t')e^{\int_{t'}^{t}[2\mathbf{t}-\mathbf{\Gamma}(t'')]dt''}.\\
\end{split}
\end{equation}

Next we study the lesser self-energy of Majorana nanowire, $\Sigma_{ik}^{<}(t,t'')=\sum_{s=L,R}[\Sigma_{ik,s}^{e<}(t,t'')+\Sigma_{ik,s}^{h<}(t,t'')]$. By the Eq. (\ref{KeldyshSE}) and making use of the definition of the free Green function and the level broadening matrix, we obtain
\begin{equation}
\begin{split}
\Sigma_{ik}^{<}(t,t'')&=i\sum_{s=L,R}\int\frac{d\varepsilon}{2\pi}e^{-i\varepsilon(t-t'')}f_{s}(\varepsilon)\\
&\{[\Gamma_{s}(\varepsilon,t,t'')]_{ik}+[\Gamma_{s}^{*}(-\varepsilon,t,t'')]_{ik}\}.\\
\end{split}
\end{equation}
With the wide-band approximation, the lesser self-energy for the wire can be further reduced to
\begin{equation}
\begin{split}
\mathbf{\Sigma}^{<}(t,t'')&=\sum_{s=L,R}\int\frac{d\varepsilon}{2\pi}e^{-i\varepsilon(t-t'')}[if_{s}(\varepsilon)]\\
&[\mathbf{\Gamma}_{s}e^{ie\int_{t''}^{t}U_{s}(t')dt'}+\mathbf{\Gamma}_{s}^{*}e^{-ie\int_{t''}^{t}U_{s}(t')dt'}].\\
\end{split}
\end{equation}

\section{useful formulae for the free Green functions}
\label{freeGF}

We first study the free lesser Green function $G_{p,s}^{0<}(t,t')=i\langle c_{p,s}^{\dag}(t')c_{p,s}(t)\rangle_{0}$. By the equation of motion method, we find that this Green function satisfies the differential equation, $\partial_{t}G_{p,s}^{0<}(t,t')=-i\xi_{p,s}(t)G_{p,s}^{0<}(t,t')$, which has the solution $G_{p,s}^{0<}(t,t')=G_{p,s}^{0<}(t',t')e^{-i\int_{t'}^{t}\xi_{p,s}(t'')dt''}$. The coefficient $G_{p,s}^{0<}(t',t')$ is just the equilibrium Fermi function for the free electron with energy $\varepsilon_{p,s}$: $G_{p,s}^{0<}(t',t')=if_{s}(\varepsilon_{p,s})=i/(e^{\beta_{s}\varepsilon_{p,s}}+1)$. Therefore, we get
\begin{equation}
\begin{split}
G_{p,s}^{0<}(t,t')&=if_{s}(\varepsilon_{p,s})e^{-i\int_{t'}^{t}\xi_{p,s}(t'')dt''},\\
\end{split}
\end{equation}
Similarly, it is easy to check that
\begin{equation}
\begin{split}
G_{p,s}^{0>}(t,t')&=-i(1-f_{s}(\varepsilon_{p,s}))e^{-i\int_{t'}^{t}\xi_{p,s}(t'')dt''},\\
\overline{G}_{p,s}^{0<}(t,t')&=i(1-f_{s}(\varepsilon_{p,s}))e^{i\int_{t'}^{t}\xi_{p,s}(t'')dt''},\\
\overline{G}_{p,s}^{0>}(t,t')&=-if_{s}(\varepsilon_{p,s})e^{i\int_{t'}^{t}\xi_{p,s}(t'')dt''}.\\
\end{split}
\end{equation}

We now calculate the retarded and advanced Green functions for the free electron \cite{Mahan1990}. By the relations, $G^{R}(t,t')=\theta(t-t')G^{>}(t,t')+\theta(t'-t)G^{<}(t,t')-G^{<}(t,t')$ and $G^{A}(t,t')=\theta(t-t')G^{>}(t,t')+\theta(t'-t)G^{<}(t,t')-G^{>}(t,t')$, and substituting the free lesser and greater Green functions into these relations, we finally arrive at
\begin{equation}
\begin{split}
G_{p,s}^{0R}(t,t')&=-i\theta(t-t')e^{-i\int_{t'}^{t}\xi_{p,s}(t'')dt''},\\
G_{p,s}^{0A}(t,t')&=i\theta(t'-t)e^{-i\int_{t'}^{t}\xi_{p,s}(t'')dt''},\\
\overline{G}_{p,s}^{0R}(t,t')&=-i\theta(t-t')e^{i\int_{t'}^{t}\xi_{p,s}(t'')dt''},\\
\overline{G}_{p,s}^{0A}(t,t')&=i\theta(t'-t)e^{i\int_{t'}^{t}\xi_{p,s}(t'')dt''}.\\
\end{split}
\end{equation}

\begin{widetext}

\section{renormalization analysis of correlation function}
\label{RGC}

Here we first give some details of the calculation of correlation  $R(r_{1}-r_{2})=\langle e^{ia\sqrt{2}\varphi(r_{1})}e^{-ia\sqrt{2}\varphi(r_{2})}\rangle_{H_{\text{nw}}}$, and then do a RG analysis for this correlation function. Substituting this correlation into the replica method Eq. (\ref{OHnw}), and expanding $D$ to the first order and $\Delta$ to the second order,
up to the first leading terms in $D$ and $\Delta$, we have $R(r_{1}-r_{2})=R_{0}+R_{\Delta}+R_{D}$, where
\begin{equation}
\begin{split}
R_{0}&=\langle e^{ia\sqrt{2}\varphi(r_{1})}e^{-ia\sqrt{2}\varphi(r_{2})}\rangle_{0},\\
R_{\Delta}&=\frac{\Delta^{2}\sin^2{\gamma_{k_{F}}}}{8(\pi\alpha u)^{2}}\sum\limits_{\epsilon=\pm1}\int d^{2}r'd^{2}r''[\langle e^{ia\sqrt{2}\varphi(r_{1})}e^{-ia\sqrt{2}\varphi(r_{2})}e^{i2\epsilon\theta(r')}e^{-i2\epsilon\theta(r'')}\rangle_{0}\\
&-\langle e^{ia\sqrt{2}\varphi(r_{1})}e^{-ia\sqrt{2}\varphi(r_{2})}\rangle_{0}\langle e^{i2\epsilon\theta(r')}e^{-i2\epsilon\theta(r'')}\rangle_{0}],\\
R_{D}&=\frac{D\cos^2{\gamma_{k_{F}}}}{8(\pi\alpha u)^{2}}\sum\limits_{\epsilon=\pm1}\int d^{2}r'd^{2}r''\delta(x'-x'')[\langle e^{ia\sqrt{2}\varphi(r_{1})}e^{-ia\sqrt{2}\varphi(r_{2})}e^{i2\epsilon\varphi(r')}e^{-i2\epsilon\varphi(r'')}\rangle_{0}\\
&-\langle e^{ia\sqrt{2}\varphi(r_{1})}e^{-ia\sqrt{2}\varphi(r_{2})}\rangle_{0}\langle e^{i2\epsilon\varphi(r')}e^{-i2\epsilon\varphi(r'')}\rangle_{0}].\\
\end{split}
\end{equation}
The average $\langle\rangle_{0}$ is performed for the Luttinger Hamiltonian $H_{\text{Lutt}}$ in Eq. (\ref{Luttingerliquid}). These correlation functions can be calculated by the following formula \cite{Giamarchi}:
\begin{equation}
\label{ephietheta}
\begin{split}
\langle\prod\limits_{j}e^{iA_{j}\phi(r_{j})}\prod\limits_{j}e^{iB_{j}\theta(s_{j})}\rangle_{0}=e^{\frac{1}{2}\sum_{i<j}A_{i}A_{j}KF_{1}(r_{i}-r_{j})}e^{\frac{1}{2}\sum_{i<j}B_{i}B_{j}K^{-1}F_{1}(s_{i}-s_{j})}e^{-\frac{1}{2}\sum_{i,j}A_{i}B_{j}F_{2}(r_{i}-s_{j})},\\
\end{split}
\end{equation}
where $r=(x,u\tau)$ and $s=(x',u\tau')$. Notice that the correlations are nonzero only when the coefficients $A_{i}$ and $B_{i}$ satisfy the neutral conditions: $\sum_{i}A_{i}=0$ and $\sum_{i}B_{i}=0$, otherwise the correlations are vanishing.
The functions
\begin{equation}
\begin{split}
F_1(r)&=\frac{1}{2}\ln{\frac{x^2+(u|\tau|+\alpha)^2}{\alpha^2}},\\
F_2(r)&=-i\text{Arg}(y_{\alpha}+ix),\\
\end{split}
\end{equation}
are the real and imaginary parts of the analytical function $\ln(y_{\alpha}-ix)$, where $y_{\alpha}=u\tau+\alpha\text{sign}(\tau)$.

For the $R_{\Delta}$ term, Using Eq. (\ref{ephietheta}), and replacing the
integration variables by $r'=R+\frac{r}{2}$ and $r''=R-\frac{r}{2}$, we have
\begin{equation}
\begin{split}
R_{\Delta}&=\frac{\Delta^{2}\sin^2{\gamma_{k_{F}}}}{4(\pi\alpha u)^{2}}\int d^{2}Rd^{2}r e^{-a^{2}KF_{1}(r_{1}-r_{2})}e^{-2K^{-1}F_{1}(r)}[a^{2}(r\cdot\nabla_{R}[F_{2}(r_{1}-R)-F_{2}(r_{2}-R)])^{2}].\\
\end{split}
\end{equation}
Since $F_1(r)$ and $F_2(r)$ are the real and imaginary parts of the analytical function $\ln(y_{\alpha}-ix)$, they obey the standard Cauchy relations: $\nabla_{X}F_{1}=i\nabla_{Y}F_{2}$, $\nabla_{Y}F_{1}=-i\nabla_{X}F_{2}$, where $R=(X,Y)$. Thus $R_{\Delta}$ can be further reduced to
\begin{equation}
\label{RDelta}
\begin{split}
R_{\Delta}&=\frac{\Delta^{2}\sin^2{\gamma_{k_{F}}}}{4(\pi\alpha u)^{2}}\int d^{2}Rd^{2}r e^{-a^{2}KF_{1}(r_{1}-r_{2})}e^{-2K^{-1}F_{1}(r)}\frac{a^{2}r^{2}}{2}\\
&\times\left[(F_{1}(r_{1}-R)-F_{1}(r_{2}-R))(\nabla_{X}^{2}+\nabla_{Y}^{2})(F_{1}(r_{1}-R)-F_{1}(r_{2}-R))\right].\\
\end{split}
\end{equation}
Note that $F_{1}(r)$ is essentially $\ln(r/\alpha)$ because the short-range cutoff $\alpha\ll r$. Therefore, one can apply the identity $(\nabla_{X}^{2}+\nabla_{Y}^{2})\log(R)=2\pi\delta(R)$ to Eq. (\ref{RDelta}), and finally find that
\begin{equation}
\begin{split}
R_{\Delta}&=-\frac{\Delta^{2}\alpha^{2}a^{2}\sin^2{\gamma_{k_{F}}}}{u^{2}}e^{-a^{2}KF_{1}(r_{1}-r_{2})}F_{1}(r_{1}-r_{2})\int_{\alpha}^{\infty}\frac{dr}{\alpha}\left(\frac{r}{\alpha}\right)^{3-2K^{-1}}.\\
\end{split}
\end{equation}
By the similar technique, we can obtain the $R_{D}$ term as follows,
\begin{equation}
\begin{split}
R_{D}&=\frac{Da^{2}K^{2}\cos^2{\gamma_{k_{F}}}}{4(\pi\alpha u)^{2}}e^{-a^{2}KF_{1}(r_{1}-r_{2})}[J_{+}I_{+}(r_{1}-r_{2})+J_{-}I_{-}(r_{1}-r_{2})],\\
\end{split}
\end{equation}
where
\begin{equation}
\label{JJII}
\begin{split}
J_{\pm}&=\int d^{2}r\delta(x)e^{-2KF_{1}(r)}(x^{2}\pm y^{2}),\\
I_{\pm}(r_{1}-r_{2})&=\int d^{2}RF_{1}(r_{1}-R)(\nabla_{X}^{2}\pm\nabla_{Y}^{2})F_{1}(R-r_{2}).\\
\end{split}
\end{equation}
Substituting $F_{1}(r)=\ln{(r/\alpha)}$ into Eq. (\ref{JJII}), we have
\begin{equation}
\begin{split}
&J_{+}=2\alpha^{3}\int\frac{dr}{\alpha}\left(\frac{r}{\alpha}\right)^{2-2K},\\
&J_{-}=-2\alpha^{3}\int\frac{dr}{\alpha}\left(\frac{r}{\alpha}\right)^{2-2K},\\
&I_{+}(r_{1}-r_{2})=2\pi F_{1}(r_{1}-r_{2}),\\
&I_{-}(r_{1}-r_{2})=\pi\cos{2\theta_{r_{1}-r_{2}}}.\\
\end{split}
\end{equation}
Finally, we obtain that the $R_{D}$ term is
\begin{equation}
\begin{split}
R_{D}&=\frac{D\alpha a^{2}K^{2}\cos^2{\gamma_{k_{F}}}}{2\pi u^{2}}e^{-a^{2}KF_{1}(r_{1}-r_{2})}[2F_{1}(r_{1}-r_{2})-\cos{2\theta_{r_{1}-r_{2}}}]\int_{\alpha}^{\infty}\frac{dr}{\alpha}\left(\frac{r}{\alpha}\right)^{2-2K},\\
\end{split}
\end{equation}
where $\theta_{r}$ is the angle between the vector $r=(x,u\tau)$ and the $x$-axis. Notice that the $\delta(x)$ term in Eq. (\ref{JJII}) makes $x$ and $u\tau$ inequivalent in $R_{D}$ term. Thus the space and time are asymmetric and have to be renormalized separately. We set
\begin{equation}
\begin{split}
F_{t}(r_{1}-r_{2})=F_{1}(r_{1}-r_{2})+\frac{t}{K}\cos{2\theta_{r_{1}-r_{2}}},\\
\end{split}
\end{equation}
where $t$ parameterizes the anisotropy between the space and time
directions, and $t=0$ in the original Hamiltonian but will be
generalized during the renormalization due to the $\delta(x)$ term.

Therefore, keeping the zeroth order term of $t$ during the renormalization, the correlation for the whole Hamiltonian should be
\begin{equation}
\label{R12}
\begin{split}
R(r_{1}-r_{2})=e^{-a^{2}KF_{t}(r_{1}-r_{2})}\bigg\{1&-a^{2}F_{1}(r_{1}-r_{2})\left[y_{\Delta}^{2}\int\frac{dr}{\alpha}\left(\frac{r}{\alpha}\right)^{3-2K^{-1}}-y_{D}K^{2}\int\frac{dr}{\alpha}\left(\frac{r}{\alpha}\right)^{2-2K}\right]\\
&-a^{2}\frac{y_{D}K^{2}}{2}\cos{2\theta_{r_{1}-r_{2}}}\int\frac{dr}{\alpha}\left(\frac{r}{\alpha}\right)^{2-2K}\bigg\},\\
\end{split}
\end{equation}
where $y_{\Delta}=\frac{\alpha\Delta\sin{\gamma_{k_{F}}}}{u}$ and
$y_{D}=\frac{\alpha D\cos^{2}{\gamma_{k_{F}}}}{\pi u^{2}}$. It is worth noting that $R(r_{1}-r_{2})$ is structurally identical to the correlation function of Luttinger Hamiltonian Eq. (\ref{Luttingerliquid}), $R_{0}(r_{1}-r_{2})=e^{-a^{2}KF_{t}(r_{1}-r_{2})}|_{t=0}$. Quantitatively, this structural similarity can be achieved by re-exponentiating Eq. (\ref{R12}), and comparing with $R_{0}(r_{1}-r_{2})$. We find that
an effective Luttinger Hamiltonian with renormalized
$K_{\text{eff}}$ and $t_{\text{eff}}$ shown below will generate the same correlation of the original Luttinger Hamiltonian (without disorder and superconductivity),
\begin{equation}
\begin{split}
K_{\text{eff}}&=K+y_{\Delta}^{2}\int_{\alpha}^{\infty}\frac{dr}{\alpha}\left(\frac{r}{\alpha}\right)^{3-2K^{-1}}-y_{D}K^{2}\int_{\alpha}^{\infty}\frac{dr}{\alpha}\left(\frac{r}{\alpha}\right)^{2-2K},\\
t_{\text{eff}}&=t+\frac{y_{D}K^{2}}{2}\int_{\alpha}^{\infty}\frac{dr}{\alpha}\left(\frac{r}{\alpha}\right)^{2-2K}.\\
\end{split}
\end{equation}
Note that generally the Luttinger parameters $K_{\text{eff}}$ and $t_{\text{eff}}$ are divergent in one dimension. However, since the Luttinger parameters
determine the correlations and thus physical properties of the
system, they should be independent of the short-range cutoff $\alpha$. It is necessary to keep the divergent Luttinger parameters as constants to preserve the physical properties of the system. Therefore, we can use the
following renormalization procedure to extract useful information
from these infinities. For $K_{\text{eff}}$, by writing the integral $\int_{\alpha}^{\infty}=\int_{\alpha}^{\alpha e^{l}}+\int_{\alpha e^{l}}^{\infty}$, integrating the first part, and rescaling the second part $\alpha e^{l}\rightarrow\alpha$, we observe that when
\begin{equation}
\begin{split}
&K(l)=K(0)+y_{\Delta}^{2}(0)\frac{e^{[4-2K^{-1}(0)]l}-1}{4-2K^{-1}(0)}-y_{D}(0)K^{2}(0)\frac{e^{[3-2K(0)]l}-1}{3-2K(0)},\\
&y_{\Delta}^{2}(l)=y_{\Delta}^{2}(0)e^{[4-2K^{-1}(0)]l},\\
&y_{D}(l)K^{2}(l)=y_{D}(0)K^{2}(0)e^{[3-2K(0)]l},\\
\end{split}
\end{equation}
$K_{\text{eff}}$ is unchanged. Sending $l$ to zero, we have
\begin{equation}
\begin{split}
&\frac{dK}{dl}=y_{\Delta}^{2}-y_{D}K^{2},\\
&\frac{dy_{\Delta}}{dl}=(2-K^{-1})y_{\Delta},\\
&\frac{dy_{D}}{dl}=2Ky_{D}^{2}-(2K-3+2K^{-1}y_{\Delta}^{2})y_{D}.\\
\end{split}
\end{equation}
Similarly, for the $t_{\text{eff}}$, we have
\begin{equation}
\begin{split}
\frac{dt}{dl}=\frac{y_{D}K^{2}}{2}.\\
\end{split}
\end{equation}
For the form of $F_{t}(r)$, a renormalization of $t$ is equivalent
to a renormalization of the velocity $u$ by
\begin{equation}
\begin{split}
-\frac{2}{K}\frac{dt}{dl}=\frac{1}{u}\frac{du}{dl}.\\
\end{split}
\end{equation}
Therefore, we have
\begin{equation}
\begin{split}
\frac{du}{dl}=-y_{D}Ku.\\
\end{split}
\end{equation}

Given a set of initial parameters, the Hamiltonian with parameters generated by the above renormalization flow equations is in the same phase. Thus we can use these renormalization flows to depict the phase diagram of the system.

\end{widetext}


%

\end{document}